\documentclass[prd, 11pt,nofootinbib, superscriptaddress, preprintnumbers,floatfix]{revtex4}

\usepackage[utf8]{inputenc}
\usepackage{amsmath,amssymb}
\usepackage{graphicx}
\usepackage{subfigure}
\usepackage{hyperref}
\usepackage{nicefrac}
\usepackage{xr}
\usepackage{footnote}
\usepackage{booktabs}
\usepackage{import}
\usepackage{braket}
\usepackage{multirow}
\usepackage{here}
\usepackage{color}
\usepackage{dsfont}
\usepackage{placeins}

\newcommand{\be}{\begin{equation}}
\newcommand{\ee}{\end{equation}}
\newcommand{\eq}{\begin{eqnarray}}
\newcommand{\en}{\end{eqnarray}}
\newcommand{\bea}{\begin{eqnarray}}
\newcommand{\eea}{\end{eqnarray}}

\newcommand{\bc}{\begin{center}}
\newcommand{\ec}{\end{center}}

\begin{document}

\title{Two- and three-body interactions in $\varphi^4$ theory from lattice simulations}
%\preprint{arXiv:XXXX.YYYYY}

\newcommand\bn{HISKP and BCTP, Rheinische Friedrich-Wilhelms Universit\"at Bonn, 53115 Bonn, Germany}
\newcommand\vlc{IFIC, CSIC-Universitat de Val\`encia, 46980 Paterna, Spain}

\author{F.~Romero-L\'opez}\affiliation{\vlc}
\author{A.~Rusetsky}\affiliation{\bn}
\author{C.~Urbach}\affiliation{\bn}

\begin{abstract}
  We calculate the one-, two- and three-particle energy levels for different lattice
  volumes in the complex $\varphi^4$ theory on the lattice.
  We argue that the exponentially suppressed finite-volume corrections for
  the two- and three-particle energy shifts can be reduced significantly by using the single
  particle mass, which includes the finite-size effects.
  We show numerically that, for a set of bare parameters, corresponding to the weak repulsive
  interaction, one can reliably extract the two- and three-particle energy shifts.
  From those, we extract the scattering length, the effective range and the effective
  three-body coupling.
  We show that the parameters, extracted from the two- and three-particle energy shifts,
  are consistent.
  Moreover, the effective three-body coupling is significantly different from zero.
\end{abstract}

\maketitle

\section{Introduction}

In the last decade, the use of the (multi-channel) L\"uscher 
equation~\cite{Luescher-torus,Rummukainen,Lage-KN,Lage-scalar,He,Sharpe,Briceno-multi,Liu,PengGuo-multi,Morningstar:2017spu} 
has become a standard tool for the extraction of the infinite-volume 
two-particle scattering phase shifts and inelasticities from the 
energy spectrum on the lattice~\cite{Helmes:2017smr, Liu:2016cba,Helmes:2015gla,Romero-Lopez:2018zyy,Wilson-pieta,Bolton:2015psa,Briceno:2016mjc,Briceno:2017qmb,Guo:2018zss}. Further, the Lellouch-L\"uscher 
formalism, as well as its generalizations to the case of the coupled 
two-particle channels, have been successfully used for the measurement of various 
matrix elements with two-particle final states, as well as for the calculation
of the resonance form factors~\cite{Lellouch:2000pv,Hansen:2012tf,Briceno:2014uqa,Briceno:2015csa,Bernard:2012bi,Agadjanov:2014kha,B-decays,Briceno:2016kkp}.
Very recently, the focus in the development of the formalism has been shifted
from two- to three-particle systems, where a significant amount of work has been done
in the course of last few years~\cite{Polejaeva:2012ut,Meissner:2014dea,Guo:2016fgl,Guo:2017ism,Briceno:2012rv,Hansen:2014eka,Hansen:2015zga,Hansen:2015zta,Hansen:2016fzj,Hansen:2016ync,Briceno:2017tce,Kreuzer:2010ti,Kreuzer:2009jp,Kreuzer:2008bi,Kreuzer:2012sr,Sharpe-recent,pang1,pang2,Meng:2017jgx,Doring:2018xxx,Mai:2017bge,Briceno:2018mlh}.
In particular, the formalism of Refs.~\cite{pang1,pang2,Doring:2018xxx}
provides a systematic fit strategy to the data, even if the latter consists
of only few data points with sizable errors (a conceptually similar approach was proposed
in Ref.~\cite{Mai:2017bge}). The output of this fit determines the
low-energy constants in the three-particle sector (three-particle force), which
can be further used, as an input, in the infinite-volume scattering equations,
in order to determine all $S$-matrix elements in the three-particle sector. It is not
immediately clear, however, whether this ambitious program can be 
realized in practice or, in other words, whether the three-body force can be
reliably extracted from the lattice data. The present study intends to explore this 
gap and test the viability of the theoretical framework. 

The strategy laid out in Refs.~\cite{pang1,pang2,Doring:2018xxx} could be
briefly summarized as follows: the interactions in the effective Lagrangian,
describing three particles, are parameterized by two sets of couplings. The 
first set of couplings describes the low-energy interactions in the two-particle
sub-systems. These can be related to the usual effective range expansion
parameters: the scattering length $a$, the effective range $r$, and so on.
The second set corresponds to the tower of the three-body local couplings,
which are multiplying the operators with an increasing number of space derivatives
acting on the fields. At lowest order, there is just a single constant
describing the contact non-derivative interaction of three particles at one
point. If the particle-dimer picture is used, these two sets of couplings
are uniquely translated into the sets describing the dimer-two-particle vertex
and the contact particle-dimer interactions. The details are given in
Refs.~\cite{pang1,pang2}.

Next, note that the couplings in the two-body sector can be 
independently fitted to the two-particle phase shift, extracted from the lattice data
by using the L\"uscher method. On the contrary, the finite-volume spectrum
in the three-particle sector, which is determined by the quantization condition
derived in Ref.~\cite{pang2}, depends on both sets of couplings, with the former already fixed from the fit to the two-particle scattering data. Thus, the fit
to the three-particle energy levels allows one to determine the first few
couplings in the three-particle (particle-dimer) sector. Using 
the values of the couplings, determined in this way,
one may then calculate all infinite volume $S$-matrix elements.

Note also that the presence of the three-body force in the equations (both
in infinite and finite volume) is crucial. These equations contain
the UV cutoff and, without the three-body force,
one could not ensure UV cutoff-independence of the observable quantities,
obtained by solving the equations. Namely, the coupling
constants should show the log-periodic dependence on the cutoff (see,
e.g.,~\cite{Bedaque:1998kg,Bedaque:1998km})\footnote{The case of a
  covariant formulation was recently investigated in
  Ref.~\cite{Gegelia}}.

The aim of our investigation is\footnote{In the context of QCD,
  the multi-pion states have been studied in Refs.~\cite{Beane:2007es,Detmold:2008fn}
in a very similar fashion.}, in particular,
 to answer to the following questions:
\begin{itemize}
\item[i)]
Is the theoretical approach, proposed in Refs.~\cite{pang1,pang2,Doring:2018xxx}, practically
viable? Could one unambiguously extract the effect from the genuinely three-body
(particle-dimer) coupling(s) from data with the present computational 
res\-so\-ur\-ces? 

\item[ii)]
What is the best fit strategy? At which volumes should the data be taken to
be most sensitive to the three-particle coupling(s)? 

\item[iii)]
Does one have all finite-volume artifacts under sufficient control in order
to extract the three-body coupling(s) with a reasonable systematic error?

\end{itemize}
In order to answer these questions we have used our framework
from the beginning to the end to analyze the data in the toy model described
by the $\varphi^4$ Lagrangian \footnote{For this model, the
    three-particle energies have also been studied in
    Ref.~\cite{Gattringer:2018ous}, however, without including the
    three-body force.}. On the one hand, this model is chosen for the
illustration 
of the fact that the three-body calculations can be systematically 
carried out even without committing extraordinary computational capabilities.
On the other hand, we use this model as an example to discuss the general issues, 
which were outlined above. 

For simplicity, we will from now on assume that the three-particle
force is described by the single non-derivative coupling. It will be seen,
{\it a posteriori,} that the accuracy of lattice data 
is not yet sufficient for extracting higher-order couplings as well.
  
It is important to realize that the role of the three-body force in the 
following situations is physically very different:
\begin{itemize}

\item[i)]
There exists a three-particle bound state: in nature this scenario
is realized e.g. in a particular three-nucleon systems (triton).
In this case, one can directly extract the pertinent three-body
coupling by extrapolating the measured bound-state energy to infinite volume.
\item[ii)]
There is no three-particle bound state: in nature this corresponds to e.g.
the final-state interactions in the $K\to 3\pi$ decay. In large volumes,
the free three-particle energy levels are displaced 
(such a case was considered, e.g., in Refs.~\cite{Huang,Wu,Tan,Beane,Hansen:2015zta,Hansen:2016fzj,Sharpe-recent}).
The displacement contains the three-particle coupling only at N$^3$LO 
in the perturbative expansion in $1/L$, where $L$ is the size of the box and, hence, it is
legitimate to ask, whether this coupling can be extracted from data at any 
precision. Further, making $L$ smaller in order to enhance the relative
contribution of the three-particle force, one is necessarily confronted
with the question about the exponentially suppressed finite volume effects.
Does one control these effects to a sufficient precision, to ensure
a clean extraction of the three-particle coupling?

\end{itemize}
We stress here once more that in nature one encounters both scenarios, so it
is useful to address them both in the model study. Whereas the extraction of the
three-particle coupling in scenario i) is relatively straightforward, the
scenario ii) should be given more scrutiny.

In particular, we would like to point out the issue of consistency of the infinite-volume and 
the finite-volume descriptions of the same system in the framework we are
considering. Suppose that scenario ii) is realized. Then, the three-body
coupling appears only at order $L^{-6}$ in the perturbative expansion of the
ground-state energy level, which starts at $O(L^{-3})$ (to be more precise,
a certain cutoff-independent combination of this coupling and the UV cutoff
should appear in the expression of the \emph{finite-volume energy}). Owing to this fact, the 
extracted value of this coupling will potentially have large errors. This could
lead to problems, if the same coupling would emerge at the leading order
in the
\emph{infinite-volume scattering amplitude.} It can be seen, however, that
this is not the case -- the pertinent contribution to the scattering amplitude is
suppressed in perturbation theory.

Finally, note that the quantization condition given in 
Ref.~\cite{pang2} determines the energy levels implicitly as the
solutions of the secular equation. {Perturbative expansion in the vicinity
of the unperturbed energy levels (both for the ground state and 
the excited states)
is possible~\cite{JWu}.
As expected,
after this expansion, the results of
Refs.~\cite{Huang,Wu,Tan,Beane,Hansen:2015zta,Hansen:2016fzj,Sharpe-recent}
for the ground state are 
reproduced\footnote{The relativistic corrections do not match, of course --
  in the approach of Ref.~\cite{pang2} the relativistic effects are not included yet. 
  It is however relatively straightforward to isolate and identify these
  effects in the final expressions.} -- there is
nothing more about the quantization condition in this case.
Therefore, in order to fit the three-particle energy levels, we shall further
use the expressions given in Ref.~\cite{Sharpe-recent}. A detailed comparison of the threshold expansion given in Ref.~\cite{Sharpe-recent} and the expansion emerging in
our approach, es well at the explicit result for the first excited three-particle state is forthcoming~\cite{JWu}.}

\section{Description of the Model \label{sec:model}}

We now turn to the description of the model which will be used in
numerical lattice simulations. We choose the complex $\varphi^4$
theory, because the charge quantum number protects the mixing of
one-, two- and three-particle states. The continuum Euclidean
Lagrangian reads 
\begin{equation}
 \mathcal{L}_E = \partial_\mu \varphi_c^\dagger \partial_\mu \varphi_c + m^2_0 |\varphi_c|^2 +\lambda_c |\varphi_c|^4,
\end{equation}
and the discretization is performed as follows:
\begin{align}
 \partial_\mu \varphi_c &\rightarrow (\varphi_c(x+a\mu) - \varphi_c(x))/a\,, \\
 \int d^4 x &\rightarrow a^4 \sum_x, \\
 a\varphi_c(x) &\rightarrow \sqrt{\kappa} \varphi_x\,,
\end{align}
and this way, the discretized action reads
\begin{align}
 S = \sum_x &\Big( -\kappa \sum_\mu(\varphi_x^* \varphi_{x+\mu} +cc) + \lambda(|\varphi_x|^2 - 1)^2 +  | \varphi_x|^2  \Big)\,,
\end{align}
with
\begin{align}
 \lambda_c = \frac{\lambda}{\kappa^2}, \quad (a m_{0})^2 =   \frac{1-2\lambda-8\kappa}{\kappa}\,. \label{eq:defpar}
\end{align}
The interpolating operators used for the one-, two- and three-particle states are:
\eq
  \hat{\mathcal{O}}_\varphi(\mathbf  x,t) = \varphi(\mathbf x,t)\,,~ 
  \hat{\mathcal{O}}_{2 \varphi}(\mathbf  x,\mathbf  y,t)
  = \varphi(\mathbf  x,t)\varphi(\mathbf  y,t)\, ,~ 
 \hat{\mathcal{O}}_{3\varphi}( \mathbf  x,\mathbf  y,\mathbf  z,t)
 = \varphi(\mathbf  x, t) \varphi(\mathbf  y, t) \varphi(\mathbf  z, t)\, ,
 \label{eq:Ops}
\en
respectively. { In principle, one must substract the vacuum expectation value (vev) to an operator before constructing the correlation function. However, the transformation
\begin{equation}
\varphi(\mathbf x,t) \rightarrow \varphi(\mathbf x,t) e^{i\pi/n},
\end{equation}
is a symmetry of the action and the operators in Eq. \ref{eq:Ops} have a vanishing vev:
\begin{equation}
\braket{\hat{\mathcal{O}}_n(\mathbf  x,t)} = \braket{\varphi(\mathbf x,t)^n} \rightarrow -\braket{\varphi(\mathbf x,t)^n} = 0.
\end{equation}
This way,} the correlation functions for the one-particle state are built as follows:
\begin{equation}
  C_1(t) = \sum_{t'} \sum_{x,y} \braket{\hat{\mathcal{O}}_\varphi(\mathbf x,t')
    \hat{\mathcal{O}}^\dagger_\varphi(\mathbf y,t+t') }\,,
\end{equation}
and, similarly, for the two- and three-particle states. That is, we extract the one-, two- and
three-particle energies in the center-of-mass (CM) frame. 

The energies can be calculated by fitting the correlation functions to their theoretical form, which can be determined by using the Transfer Matrix formalism:
\begin{equation}
  C_i(t) = \text{Tr }(e^{-\hat H(T-t)}\hat{\mathcal{O}}_i(0)e^{-\hat Ht}
  \hat{\mathcal{O}}_i^\dagger(0)) / \mathcal{Z}\, ,\quad\quad
  \hat{\mathcal{O}}_i=\hat{\mathcal{O}}_\varphi\, ,\,\hat{\mathcal{O}}_{2\varphi}\, ,\,
    \hat{\mathcal{O}}_{3\varphi}\, ,
\end{equation}
and calculating explicitly in terms of all possible states:
\begin{align}
 \begin{split}
  C_i(t) &= \sum_{n,m} |\braket{n|\hat{\mathcal{O}}_i|m} |^2 e^{-E_n(T-t)} e^{-E_mt} /\mathcal{Z} \\
  &= \sum_{n,m} |\braket{n|\hat{\mathcal{O}}_i|m} |^2 e^{-(E_m+E_n)T/2} \cosh((E_m-E_n)(t-T/2))/\mathcal{Z}.
 \end{split}
\end{align}
This way, the different correlation functions look like:
\begin{align}
  C_1(t) &= |A_1|^2 e^{-M \frac{T}{2}} \cosh{M(t-T/2)} ,\label{eq:C1}\\
C_2(t) &= |A_2|^2 e^{-E_2 \frac{T}{2}} \cosh{E_2(t-T/2)} + |A_{1\to 1}|^2 e^{-MT}, \label{eq:C2}\\
C_3(t) &= |A_3|^2 e^{-E_3 \frac{T}{2}} \cosh{E_3(t-T/2)} + |A_{1\to 2}|^2 e^{-(E_2+M) \frac{T}{2}} \cosh{(E_2-M)(t-T/2)}\label{eq:C3}\,.
\end{align}
The additional terms in $C_2$ and $C_3$ come from thermal pollution, {i.e., antiparticles propagating backwards in time and crossing the periodic temporal boundary. They are relevant}, since
the matrix elements 
$A_{1\to 1} \propto\braket{\varphi^\dagger | \hat{ \mathcal{O}}_{2\varphi}  | \varphi}$ and
$A_{1\to 2} \propto\braket{\varphi^\dagger | \hat{ \mathcal{O}}_{3\varphi}  | \varphi\varphi}$
are different from zero and, in general, they are of the same order of magnitude
as $A_2$ and $A_3$, respectively.
In contrast to $C_2$, the pollution term in $C_3$ is time-dependent.
Both of these additional terms vanish in the limit of $T\to\infty$ 
but, for finite $T$, they have to be taken into account in the analysis.
For $C_2$, this can be done by using the discrete derivative of the correlation function,
the so-called shifted correlation function:
\begin{equation}\label{eq:shifted}
  \tilde{C}_2(t) = C_2(t)-C_2(t+1)\,.
\end{equation}
This has also advantage in terms of less correlation among the different time
slices~\cite{Ottnad:2017bjt}.
Using Eq.~(\ref{eq:shifted}) turns the functional
dependence on $t$ in $C_2(t)$ into $\sinh$.
For $C_3$, we take the time-dependent pollution into account explicitly in our fitting.
In doing so, we are using the single particle mass $M$ and the two particle
energy $E_2$ as an input, determined from $C_1$ and $C_2$, respectively.

{Our ensembles are generated using the Metropolis algorithm with simultaneous updates in the even and odd parts of the lattice.} We choose $m^2_0 = -4.9$ and $\lambda_c=10.0$
implying $\lambda=0.253308$ and $\kappa= 0.159156$. 
We calculate the spectra in 15 different volumes, $L=[5,18]$ and $L=20$ with $T=24$
and two volumes $L=14, \ 24$ with $T=48$.
The number of independent configurations is always in the range 7500-100000 and the errors are calculated through jackknife resampling. A summary of all ensembles can be found in Table \ref{tab:spectrum}. For the reasons given in the previous paragraph, 
we fit the corresponding $\sinh$ functional form to our data for the correlation
functions $\tilde{C}_1$, $\tilde{C}_2$ and $\tilde{C}_3$. { The excited states are expected to be strongly suppressed because {{we use charged operators and because of}} the small value of the renormalized coupling $\lambda_c$. In practice, there seems to be no sign 
of excited states in the correlation functions, and thus the fits are performed to all time slices. }
This can be seen in the effective mass ($m_\mathrm{eff}$) plots
compiled in Appendix \ref{app:meff}, where $m_\mathrm{eff}$ is defined through:
\begin{equation}
  \frac{\tilde{C}_1(t+1)}{\tilde{C}_1(t)} =
  \frac{\sinh(m_\mathrm{eff}(t+1-\frac{T}{2}))}{\sinh(m_\mathrm{eff}(t-\frac{T}{2}))}\, .
\end{equation}
For $C_2$, the relation looks identical. For $C_3$, the functional form is more complicated, but $m_\mathrm{eff}$ can be defined likewise.

\section{The spectrum in finite volume\label{sec:theory}}

In the $\varphi^4$ theory in the symmetric phase, which is 
studied in the present paper, the vertices with the
odd number of the field $\varphi$ are barred. 
This situation resembles the one in chiral perturbation theory (ChPT) and we can straightforwardly use
the perturbative expression, obtained in Ref.~\cite{GL} (see also Ref.~\cite{Colangelo}),
to describe the volume-dependence of the single-particle mass. This expression
has the following form
\eq\label{eq:ML1}
M_L-M=\mbox{const}\,\frac{K_1(ML)}{ML}\, ,
\en
where $M=M_\infty$ and $K_1(z)$ denotes the modified Bessel function of first kind.
Taking into account the asymptotic expression of the Bessel function for a large
value of the argument, we get
\eq\label{eq:ML2}
M_L-M=\mbox{const}\,\frac{\exp(-ML)}{(ML)^{3/2}}\, .
\en
The threshold expansion of the two- and three-particle energies can be taken
from the literature. For convenience, we generally stick to the notations used
Ref.~\cite{Sharpe-recent} (we use lattice units in the formulae below).
The two-body energy shift $\Delta E_2$ reads:
\eq\label{eq:E2}
\Delta E_2=\frac{4\pi a}{ML^3}\,\biggl\{1+c_1\biggl(\frac{a}{\pi L}\biggr)
+c_2\biggl(\frac{a}{\pi L}\biggr)^2
+c_3\biggl(\frac{a}{\pi L}\biggr)^3
+\frac{2\pi ra^2}{L^3}-\frac{\pi a}{M^2L^3}\biggr\}
+O(L^{-7})\, ,
\en
where $a$ and $r$ denote the two-body scattering length and the effective radius, respectively, and $c_1,c_2,c_3$ are known numerical constants. The last term is the
relativistic correction, which vanishes at $M\to\infty$.

The three-body energy shift of the ground-state level $\Delta E_3$ takes the form:
\begin{align}
  \begin{split}
    \Delta E_3 =&  \frac{12 \pi a}{M L^3} \biggl\{ 1 + d_1 \biggl(\frac{a}{\pi L}\biggr)
    + d_2 \left(\frac{a}{\pi L} \right)^2
    + \frac{3\pi a}{ M^2L^3}  + \frac{6\pi r a^2}{L^3} \\
    &+ \left(\frac{a}{\pi L} \right)^3 d_3 \log \frac{M L}{2\pi}   \biggr\}- \frac{D}{48 M^3 L^6}  + O(L^{-7})\,, \label{eq:E3}
\end{split}
\end{align}
where $d_1,d_2,d_3$ are the known numerical constants and the parameter $D$
is a sum of different terms including the
three-body coupling constant we are looking for
(or, the divergence-free three-body scattering amplitude ${\cal M}_{3,thr}$
in the notations of Ref.~\cite{Sharpe-recent}).
The individual terms in this sum (including the three-body coupling)
depend on the ultraviolet cutoff -- only the sum of all terms in
$D$ is cutoff-independent\footnote{Note that, in Ref.~\cite{Detmold:2008fn}, the
  dependence of the ultraviolet scale in the three-body coupling was eliminated by
  defining a volume-dependent coupling constant, i.e., the scale dependence was
  effectively traded for the $L$-dependence.}.
In addition, the quantity $D$ depends on the choice of the
scale in the logarithm, which is present in Eq.~(\ref{eq:E3}) -- there, it is chosen to be
equal to $M$, which is the only available natural scale in the theory. Choosing any
other scale around this natural scale leads to an additive contribution to $D$.
Consequently, our goal can be re-formulated as follows: we want to show that
$D$ is clearly non-zero beyond the statistical uncertainty, when the scale in the 
logarithm is of order $M$. This fact would be interpreted as a detection of the
contribution of the three-particle force, which also emerges at $O(L^{-6})$. Any further
refinement of the argument seems not to be possible, because the individual
contributions are scale- and cutoff-dependent.

Before fitting the above formulae for the energy shift to the lattice data,
the following important question is in order. As we see, in order to reliably
extract the coefficient of order $L^{-6}$, we have to go to not so large $L$. In this case,
the {\em exponentially suppressed} corrections in $L$, which were neglected
in Eqs.~(\ref{eq:E2}), (\ref{eq:E3}), might become non-negligible (and, as we
shall see, they really do). Then, what is the systematic error imposed on 
the extracted value of the coefficient of order $L^{-6}$ by neglecting such terms?
We shall show that one may treat the leading exponential corrections in a relatively simple
fashion, greatly improving the accuracy of the Eqs.~(\ref{eq:E2}) and (\ref{eq:E3}).

In the infinite volume, the two- and three-particle thresholds are located
at the energies $2M$ and $3M$, respectively. In a finite volume, this shifts
to $2M_L$ and $3M_L$. This exponentially suppressed shift, 
albeit not large, may become statistically relevant, if low $L$ values 
are included in the fit. Below we shall argue that, up to next-to-leading
order in perturbation theory, using $\Delta E_2=E_2(L)-2M_L$
and $\Delta E_3=E_3(L)-3M_L$ already captures the bulk of these exponentially
suppressed corrections -- the remaining terms should be suppressed by two
powers of $L$ and by one power of the coupling constant $\lambda_c$
with respect to the leading correction\footnote{It is conceivable that the argument
  can be extended to all orders in perturbation theory -- at least, we were not able
  to find an example of a diagram in higher orders that would invalidate it. A full proof,
  however, seems to be a rather challenging affair. Since we are dealing here with
  the perturbative case, where the coupling constant $\lambda_c$ is small, we were tempted to restrict
  ourselves to a statement, which is valid up to the next-to-leading order in
  $\lambda_c$ only.}.

In order to study the exponentially suppressed terms, one has to abandon the
non-relativistic effective field theory, which provides a very
comfortable framework to derive equations like (\ref{eq:E2}) or
(\ref{eq:E3}), and turn to the relativistic description of the system.
In the following, we work in Minkowski space and consider the
two-particle case first. It is convenient to start with the
Bethe-Salpeter equation for the two-body scattering amplitude $T$ in
the center-of-mass (CM) frame.
In the infinite volume, the equation takes the form\footnote{We
  implicitly assume that the ultraviolet divergences are cured, e.g.,
  by introducing some cutoff in the integrals. These 
divergences are not relevant in the discussion of the finite-volume effects.}:
\eq\label{eq:BS}
T(p,p')=K(p,p')+\frac{1}{2}\,\int \frac{d^4k}{(2\pi)^4i}\,
K(p,k)G_2(k)T(k,p')\, .
\en
Here, the 4-momenta $p,k,p'$ denote the relative momenta of the two-particle system
with total momentum $P=(E,{\bf 0})$. Further, $K(p,p')$ is the kernel of the
Bethe-Salpeter equation (the sum of all two-particle irreducible diagrams), and
$G_2(k)=G(\frac{1}{2}\,P+k)G(\frac{1}{2}\,P-k)$ is the product of two dressed
one-particle propagators, including all self-energy insertions. To the order we are
working, however, these self-energy insertions do not contribute and $G$ can be
replaced by the free relativistic propagator.

In a finite volume with periodic boundary conditions,
the counterpart of Eq.~(\ref{eq:BS}) is written as:
\eq
T_L(p,p')=K_L(p,p')+\frac{1}{2}\,\int \frac{dk_0}{2\pi i}\,\frac{1}{L^3}\,\sum_{\bf k}
K_L(p,k)G_{2L}(k)T_L(k,p')\, ,\quad\quad {\bf k}=\frac{2\pi}{L}\,{\bf n}\, ,\quad
{\bf n}\in\mathbb{Z}^3\, ,
\en
where $L$ stands for the spatial size of the (cubic) box (the temporal extent of the box
is assumed infinite). The three-momenta ${\bf p},{\bf p}'$ are discretized as well.
Further, in analogy with Ref.~\cite{Luescher-1}, we single out the positive-energy
contribution in the two-particle propagator by writing
\eq\label{eq:L1}
G_{2L}(k)=\frac{1}{(2w_L({\bf k}))^2}\,
\frac{2\pi i\delta(k_0)}{2w_L({\bf k})-E}+\hat G_{2L}(k)\, ,\quad\quad
w_L({\bf k})=\sqrt{M_L^2+{\bf k}^2}\, .
\en
In the above equation, $\hat G_{2L}(k)$ contains in general the
contributions from the negative-energy states (anti-particles), as well
as the contributions coming from the many-particle intermediate states
in the spectral representation. Defining further 
\eq\label{eq:L2}
\hat K_L(p,p')=K_L(p,p')+\frac{1}{2}\,\int\frac{dk_0}{2\pi i}\,\frac{1}{L^3}\,\sum_{\bf k}
K_L(p,k)\hat G_{2L}(k)\hat K_L(k,p')\, ,
\en
and
\eq\label{eq:L3}
T_L({\bf p},{\bf p}')=T_L(p,p')\biggr|_{p_0=p'_0=0}\, ,\quad\quad
\hat K_L({\bf p},{\bf p}')=\hat K_L(p,p')\biggr|_{p_0=p'_0=0}\, ,
\en
we get:
\eq\label{eq:L4}
T_L({\bf p},{\bf p}')=\hat K_L({\bf p},{\bf p}')+\frac{1}{2}\,\frac{1}{L^3}\,\sum_{\bf k}
\hat K_L({\bf p},{\bf k})\frac{1}{(2w_L({\bf k}))^2}\,
\frac{1}{2w_L({\bf k})-E}\,T_L({\bf k},{\bf p}')\, .
\en
If one now singles out the term with ${\bf k}=0$, it is possible to define
\eq\label{eq:L5}
\hat T_L({\bf p},{\bf p}')=\hat K_L({\bf p},{\bf p}')+\frac{1}{2}\,\frac{1}{L^3}\,\sum_{{\bf k}\neq{\bf 0}}
\hat K_L({\bf p},{\bf k})\frac{1}{(2w_L({\bf k}))^2}\,
\frac{1}{2w_L({\bf k})-E}\,\hat T_L({\bf k},{\bf p}')\, .
\en
Then, it can be shown that there exists an algebraic relation between $T_L$ and $\hat T_L$:
\eq\label{eq:L6}
T_L({\bf p},{\bf p}')=\hat T_L({\bf p},{\bf p}')
+\frac{\hat T_L({\bf p},{\bf 0})\hat T_L({\bf 0},{\bf p}')}
{2L^3(2M_L)^2\biggl(2M_L-E-\dfrac{1}{2L^3(2M_L)^2}\,\hat T_L({\bf 0},{\bf 0})\biggr)}\, .
\en
From the above expression, it is seen that the exact
energy shift of the ground state
is determined from the equation
\eq\label{eq:L7}
2M_L-E-\frac{1}{2L^3(2M_L)^2}\,\hat T_L({\bf 0},{\bf 0})=0\, .
\en
Note that the quantity $\hat T_L({\bf 0},{\bf 0})$ itself depends on the energy $E=E_2(L)$, so
the solution of the above equation can be written down perturbatively, expanding order by order
in the energy shift $(E-2M_L)$.

\begin{figure}[t]
\begin{center}
\includegraphics*[width=4.cm]{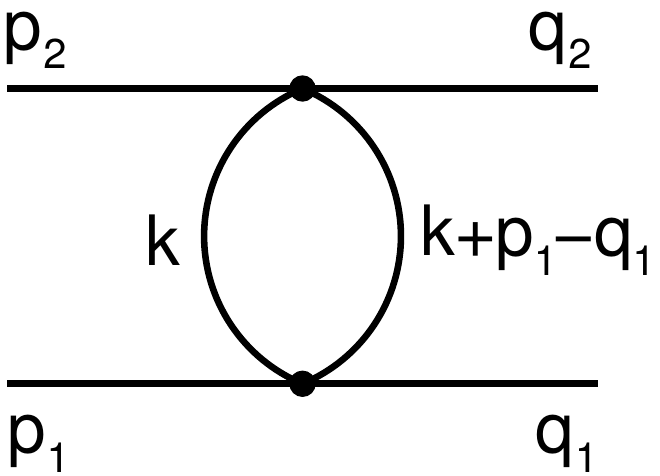}
\caption{The lowest-order term in $K^t_L$, which contains exponentially suppressed contributions.}\label{fig:tchannel}
\end{center}
\end{figure}

Let us now find the solution of this equation up to second order in the coupling
constant $\lambda_c$. We shall be interested in the exponentially suppressed terms only,
since the power-suppressed terms will obviously coincide with those in the L\"uscher
equation. At lowest order in $\lambda_c$, we get $\hat T_L({\bf 0},{\bf 0})=-24\lambda_c=\mbox{const}$, so there are no exponentially suppressed terms at all. These appear first
at order $\lambda_c^2$. The Bethe-Salpeter kernel at this order is given by
\eq
K_L(p,p')=-(24\lambda_c)+\frac{1}{2}\,(24\lambda_c)^2(K^t_L(p,p')+K^u_L(p,p'))\, ,
\en
where $K^{t,u}_L$ are the $t,u$-channel one-loop diagrams of the type shown in
Fig.~\ref{fig:tchannel}. The Feynman ``integral'' in a finite volume, which corresponds
to this diagram, is given by\footnote{We have replaced $M_L$ by $M$ everywhere, since
  the difference is exponentially suppressed and does not matter at the order we are working.}
\eq\label{eq:JL}
K^t_L(p,p')=\frac{1}{L^3}\,\int\frac{dk_0}{2\pi i}\sum_{\bf k} \frac{1}{(M^2-k_0^2+{\bf k}^2-i0)^2}=K^t_\infty(p,p')+\mbox{const}\frac{\exp(-ML)}{L^{1/2}}\, .
\en
This result can be most easily obtained by noting that Eqs.~(\ref{eq:ML1}), (\ref{eq:ML2})
are nothing but the contribution from the tadpole graph to the one-particle self energy.
Then, the leading asymptotic behavior in Eq.~(\ref{eq:JL}) is obtained by
differentiating the r.h.s. of Eq.~(\ref{eq:ML2}) with respect to $M$.

Further, using Eqs.~(\ref{eq:L2})-(\ref{eq:L5}), one obtains:
\eq
\hat T_L({\bf 0},{\bf 0})=K_L(0,0)+\frac{1}{2}\,(24\lambda_c)^2(I_L^{(1)}+I_L^{(2)})
+O(\lambda_c^3)\, ,
\en
where
\eq
I_L^{(1)}=\frac{1}{L^3}\,\sum_{\bf k}\frac{1}{(2w_L({\bf k}))^2}
\frac{1}{2w_L({\bf k})+E}\, ,\quad\quad
I_L^{(2)}=\frac{1}{L^3}\,\sum_{{\bf k}\neq{\bf 0}}\frac{1}{(2w_L({\bf k}))^2}
\frac{1}{2w_L({\bf k})-E}\, .
\en
Using Poisson's summation formula, one can single out the leading exponential
correction to the infinite-volume result:
\eq
I_L^{(1)}=\sum_{{\bf j}\in\mathbb{Z}^3}\int\frac{d^3{\bf k}}{(2\pi)^3}\,
\frac{e^{iL{\bf k}{\bf j}}}{(2w_L({\bf k}))^2(2w_L({\bf k})+E)}
=I^{(1)}_\infty+6\int\frac{d^3{\bf k}}{(2\pi)^3}\,
\frac{e^{iLk_1}}{(2w_L({\bf k}))^2(2w_L({\bf k})+E)}+\cdots\, ,
\nonumber\\
\en
where the ellipses stand for the more suppressed terms. Shifting now the integration
contour in $k_1$ into the complex plane: $k_1\to k_1+iM$ (we again replace $M_L\to M$),
and rescaling the integration variables according to $k_1\to k_1/L,~k_{2,3}\to k_{2,3}/\sqrt{L}$,
we obtain:
\eq
I_L^{(1)}&=&I^{(1)}_\infty
+\frac{3e^{-ML}}{2L}\,\int\frac{dk_1d^2{\bf k}_\perp}{(2\pi)^3}\,
\frac{e^{ik_1}}{\biggl(2iMk_1+{\bf k}_\perp^2+\frac{k_1^2}{L}\biggr)
  \biggl(2\sqrt{\frac{2iMk_1+{\bf k}_\perp^2}{L}+\frac{k_1^2}{L^2}}+E\biggr)}
\nonumber\\[2mm]
&\to& I^{(1)}_\infty+\frac{3e^{-ML}}{2EL}\,\int\frac{dk_1d^2{\bf k}_\perp}{(2\pi)^3}\,
\frac{e^{ik_1}}{2iMk_1+{\bf k}_\perp^2}
=I^{(1)}_\infty+\frac{3e^{-ML}}{4EML}\,\int\frac{d^2{\bf k}_\perp}{(2\pi)^2}\,
\exp\biggl(-\frac{{\bf k}_\perp^2}{2M}\biggr)
\nonumber\\[2mm]
&=&\int\frac{d^3{\bf k}}{(2\pi)^3}\,\frac{1}{(2w({\bf k}))^2}
\frac{1}{2w({\bf k})+E}+O\biggl(\frac{e^{-ML}}{L}\biggr)\, .
\en
Here, $w({\bf k})=\sqrt{M^2+{\bf k}^2}$ and ${\bf k}_\perp=(k_2,k_3)$. It is seen that the leading
exponentially suppressed terms here have an extra suppression factor $L^{-1/2}$,
as compared to the $t,u$-channel contributions.

The calculation of $I_L^{(1)}$ proceeds analogously, with the difference that the denominator
is now singular -- so, one has to carry out subtractions in the numerator first, in order to be able
to use Poisson's formula. The final result is given below:
\eq
I_L^{(2)}&=&P.V.\int\frac{d^3{\bf k}}{(2\pi)^3}\,\frac{1}{(2w({\bf k}))^2}
\frac{1}{2w({\bf k})-E}
+\frac{1}{2E}\biggl(\frac{1}{L^3}\sum_{{\bf k}\neq{\bf 0}}\frac{1}{{\bf k}^2-q_0^2}
-P.V.\int\frac{d^3{\bf k}}{(2\pi)^3}\,\frac{1}{{\bf k}^2-q_0^2}\biggr)
\nonumber\\[2mm]
&+&\frac{1}{4EM^2L^3}\frac{E+4M}{E+2M}+O\biggl(\frac{e^{-ML}}{L}\biggr)\, ,
\en
where $q_0^2=E^2/4-M^2$.
The last two terms in the above expression give the familiar power-suppressed
contributions to the L\"uscher formula, which are already contained in Eq.~(\ref{eq:E2})
(in particular, the last term is responsible for the relativistic correction).
Putting things together, we finally conclude that the leading exponentially suppressed
corrections in $\hat T_L({\bf 0},{\bf 0})/L^3$ are of order
$\lambda_c^2e^{-ML}/L^{7/2}$, whereas in the mass $M_L$
these contribute at order $\lambda_ce^{-ML}/L^{3/2}$. Thus, calculating the
shift $\Delta E_2=E_2(L)-2M_L$ instead of $E_2(L)-2M$ one gains the suppression factor
$\lambda_c/L^2$ for the exponential contributions at no extra cost.

\begin{figure}[t]
\begin{center}
\includegraphics*[width=3.5cm]{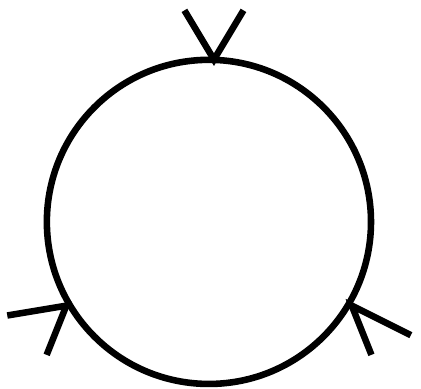}
\caption{The leading contribution to the three-particle coupling constant of the effective
  theory}\label{fig:6particle}
\end{center}
\end{figure}

Considering the three-particle energy we remark that the above result for the two-particle
energy shift could be interpreted in terms of the non-relativistic effective Lagrangians,
with the coupling constants having exponentially suppressed contributions. For
example, $M_L=M+O(e^{-ML}/L^{3/2})$, $a_L=a+O(e^{-ML}/L^{1/2})$, and so on.
One can use these volume-dependent constants in Eq.~(\ref{eq:E3}), in order to
find the leading exponential terms. Further, in this language,
the leading exponential correction
to the three-particle coupling constant arises from the diagram, shown in
Fig.~\ref{fig:6particle}. In analogy with Eq.~(\ref{eq:JL}), it is straightforward to
derive that the leading exponential corrections coming from this diagram
are suppressed by the factor $e^{-ML}L^{1/2}$. Using this information in Eq.~(\ref{eq:E3}),
we arrive at the same conclusion as in the two-particle case, namely, that replacing
$E_3(L)-3M$ by $E_3(L)-3M_L$, one gains the suppression factor $\lambda_c/L^2$
for the exponential contributions.

\section{Numerical Results}

\begin{figure}
  \centering
  \includegraphics[width=.7\linewidth]{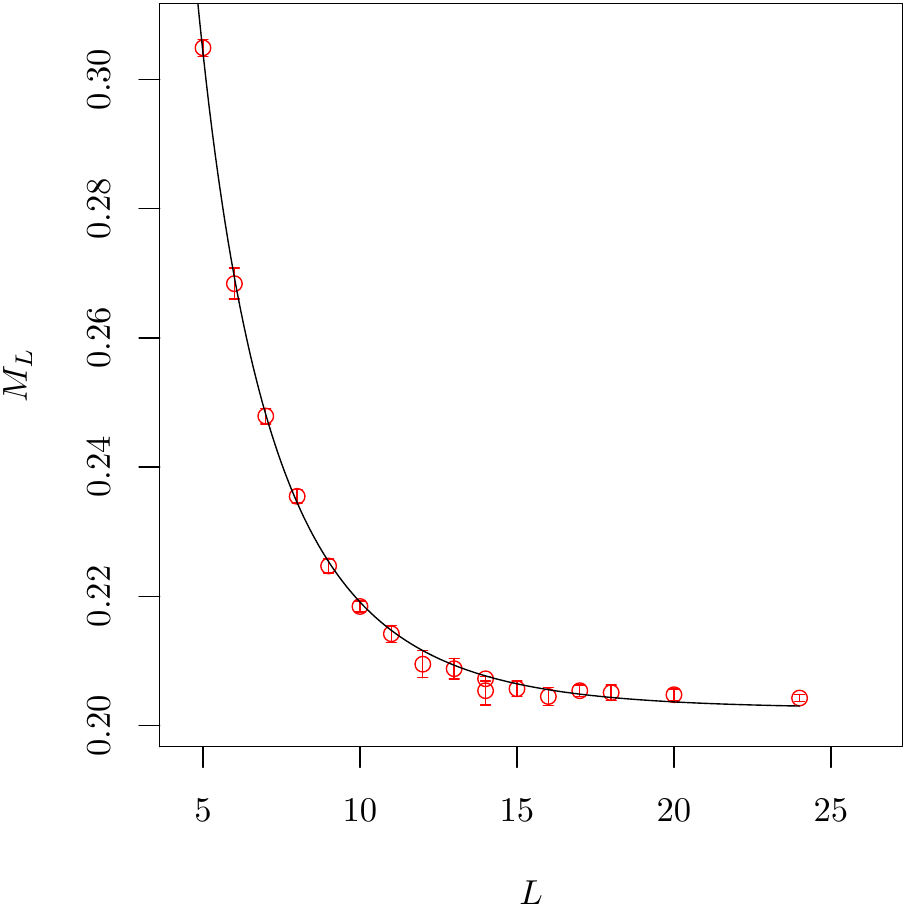}
  \caption{$M_L$ as a function of $L$.
    The solid line represents a fit of Eq.~(\ref{eq:ML1}) to the data in the fit range $[4,24]$.}
  \label{fig:MLfit}
\end{figure}

We now address the analysis of the lattice spectrum in our model,
which is shown in Table \ref{tab:spectrum} in Appendix \ref{app:spectrum}.
Moreover,  as an illustration we also show in Appendix
\ref{app:meff} the plots for the
effective mass and the correlation function for the case $L=18$.

First, we study the volume dependence of the single particle mass, which is given by
Eq.~(\ref{eq:ML1}).
The results are range-independent and seem to describe the data very well.
This is immediately seen in Fig.~\ref{fig:MLfit}, which displays the fit of Eq.~(\ref{eq:ML1})
to the data on $M_L$ in the whole range of volumes from $L_\mathrm{min}=4$ to
$L_\mathrm{max}=24$. The fit yields
\begin{equation}
M_\infty = 0.2027(2), \ \ \ \chi^2/\mathrm{d.o.f.}= 15.17/16 , \ \ \ p\,\text{-value}= 0.44\,,
\end{equation}
indicating a very good description of the data.
The best fit parameters for the fits with other fit ranges are compiled in Appendix \ref{app:allfits}.

\begin{figure}
  \centering
  \subfigure{\includegraphics[width=.49\linewidth]{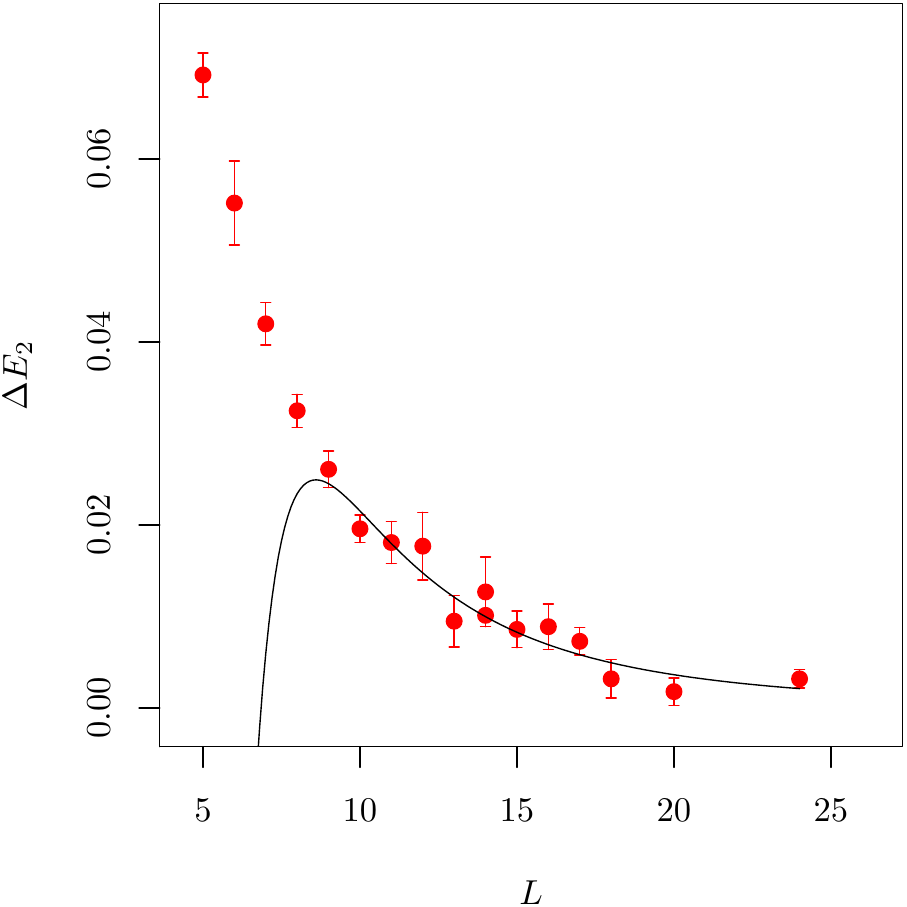}}
  \subfigure{\includegraphics[width=.49\linewidth]{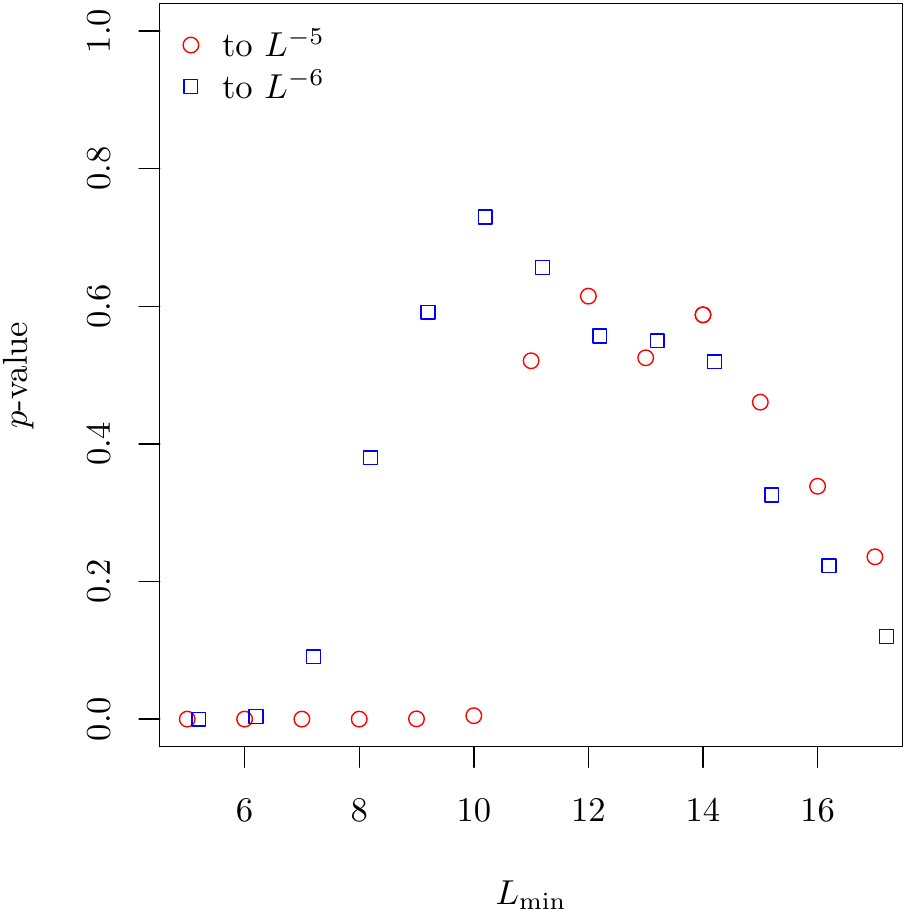}}
  \caption{Left: fit of Eq.~(\ref{eq:E2}) to the data for $\Delta E_2$ in the $L$-range $[9,24]$ including $a$ and $r$ as fit parameters.
    Right: $p$-value of the fit of Eq.~(\ref{eq:E2}) to the data for $\Delta E_2$ as a function of $L_\mathrm{min}$ with a constant $L_\mathrm{max} = 24$.}
  \label{fig:DeltaE2-1}
\end{figure}

Next, we consider the two-particle energy shift $\Delta E_2(L)$, for which data
from $L=4$ to $L=24$ are available.
We fit Eq.~(\ref{eq:E2}) to the data, varying the lower end of the fit range in
$L$ and the number of the fit parameters.
The upper end in the fit range is kept fixed at $L_\mathrm{max}=24$.
For large $L_\mathrm{min}$, one expects that only the scattering length is important and
that in Eq.~(\ref{eq:E2}) one can neglect all terms of order $L^{-6}$ or smaller. 
Adding the effective range $r$ as a fit parameter,
it should be possible to include smaller values of $L$ in the fit,
while reproducing the results of the fit with only the scattering length as a free parameter.

In the left panel of Fig.~\ref{fig:DeltaE2-1}, we show an example of a fit of
Eq.~(\ref{eq:E2}) to the data including all terms up to order $L^{-6}$, i.e.
with $a$ and $r$ as fit parameters.
$L_\mathrm{min}=9$ was chosen for this particular fit.
In the right panel of that figure, we show the $p$-value of the fit of
Eq.~(\ref{eq:E2}) to the data as a function of $L_\mathrm{min}$.
Red circles correspond to the one-parameter fit up to order $L^{-5}$,
and blue squares to the two-parameter fit up to order $L^{-6}$.
For the former, we observe the $p$-values around $0.5$ from $L_\mathrm{min}=11$
and larger, for the latter from around $L_\mathrm{min}=8$.

\begin{figure}
  \centering
  \subfigure{\includegraphics[width=.49\linewidth]{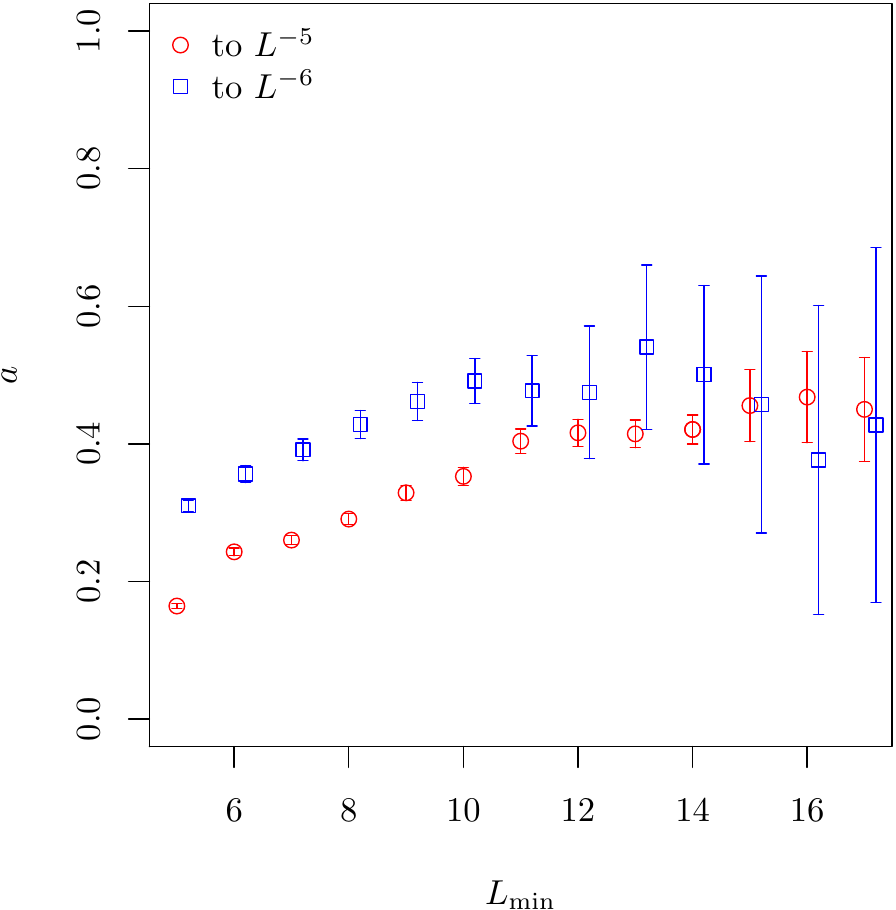}}
  \subfigure{\includegraphics[width=.49\linewidth]{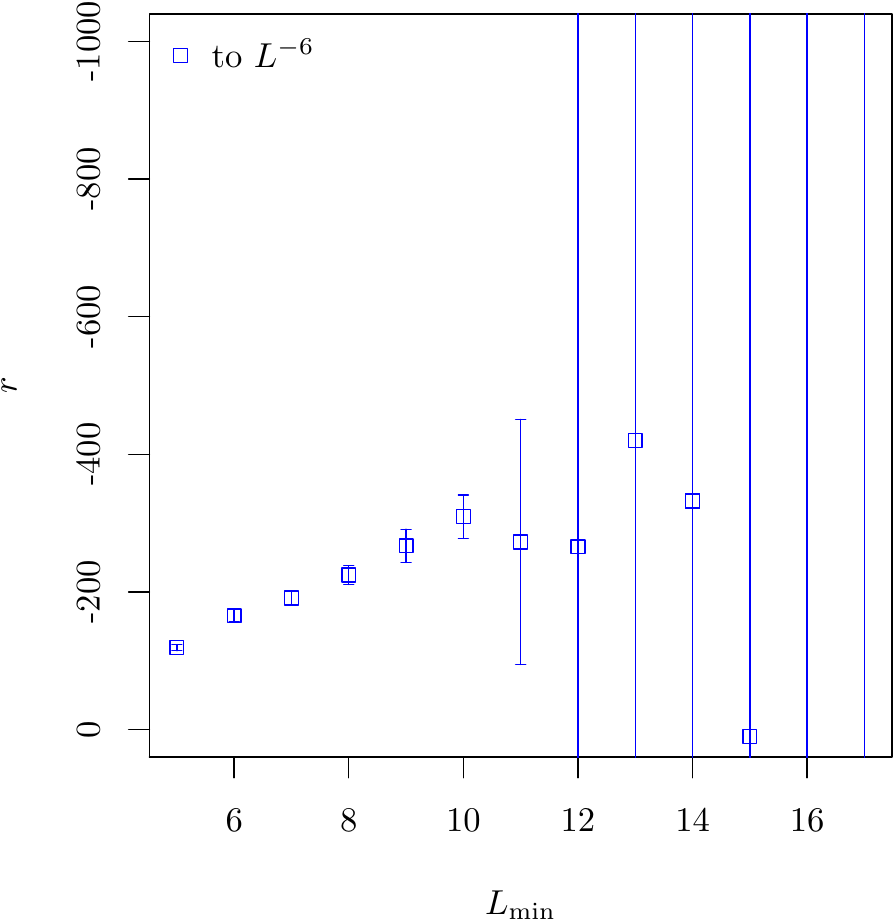}}
  \caption{The scattering length $a$ (left) and the effective range $r$ (right) determined by a fit of Eq.~(\ref{eq:E2}) to the data for $\Delta E_2$ as functions of $L_\mathrm{min}$ with fixed $L_\mathrm{max}=24$.}
  \label{fig:DeltaE2-2}
\end{figure}

From these $L_\mathrm{min}$ values on, we also observe stable values for the fit parameters $a$ and $r$ within errors.
This is shown in Fig.~\ref{fig:DeltaE2-2}, where we plot $a$ in the
left panel and $r$ in the right panel, both as functions of $L_\mathrm{min}$.
In addition, as one infers from the left panel, the values for $a$ agree within errors
between the single- and the two-parameter fits from the aforementioned $L_\mathrm{min}$-values onward.
As a result, we quote here the best fit parameters from the two-parameter fit with $L_\mathrm{min}=9$, which can be found in Table~\ref{tab:selected}.

Further, each data point for $\Delta E_2$ can be translated into one phase shift at a certain value of the scattering momentum.
For that, we calculate the S-wave phase shift as~\cite{Luescher-torus}
\begin{equation}
  \label{eq:delta}
  \cot \delta = \frac{Z_{00}(1,q^2)}{\pi^{3/2} q}\,,
\end{equation}
where $Z_{00}(1,q^2)$ is the generalized L{\"u}scher zeta-function, $q=\frac{Lk}{2\pi}$ and\footnote{{At this level, it is not important to use the lattice or continuum dispersion relation, since the momentum $k$ is small.}} $E_{2} = 2\sqrt{k^2 + M_L^2}$.
Note that we used here the mass $M_L$, taking into account the discussion in
Section~\ref{sec:theory}.
The results can be seen from Table \ref{tab:delta} in Appendix \ref{app:delta}.
The phase shift obeys the inequality $|\delta| < 2\, ^\circ$, guaranteeing
that with the current model parameters we are
in the perturbative regime.
The smallness of the phase shift also supports the validity of the arguments in Section \ref{sec:theory}.

\begin{figure}[H]
  \centering
  \includegraphics[width=.7\linewidth]{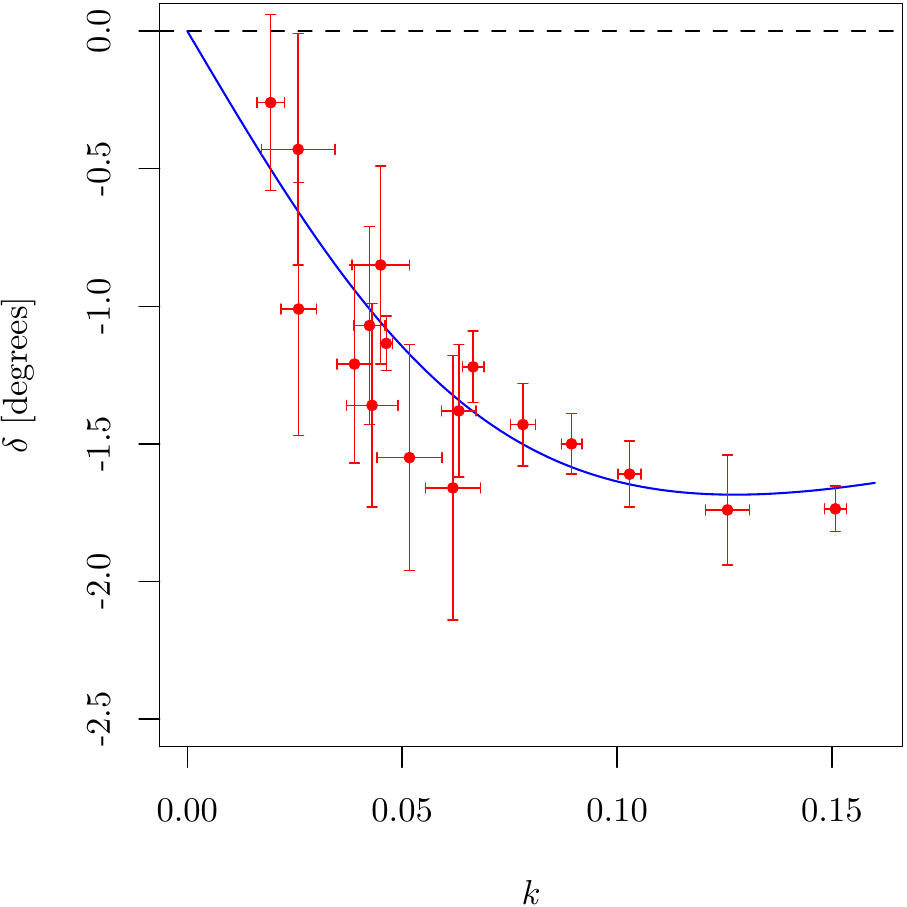}
  \caption{Phase shift as a function of the scattering
    momentum $k$ in lattice units.
    The expected curve, given by the parameters from the best fit to $\Delta E_2$,
    is shown as a solid line.}
  \label{fig:delta}
\end{figure}

In Fig.~\ref{fig:delta},
the phase shift is shown as a function of the scattering momentum $k$.
We are now able to check, whether the parameters obtained from the fit to
$\Delta E_2$ also describe the phase shift data properly.
For this purpose, we use the effective range expansion:
\begin{equation}
  \label{eq:ere}
  k \cot \delta = -\frac{1}{a} + \frac{1}{2} r k^2 + O(k^4)
\end{equation}
with the best fit parameters compiled in Table~\ref{tab:selected}.
In Figure \ref{fig:delta}, we show the phase shift (in degrees),
determined by using Eq.~(\ref{eq:delta}),
as a function of the scattering momentum $k$ (in lattice units).
In addition, we show $\delta(k)$, determined from Eq.~(\ref{eq:ere})
by using the best fit parameters from the fit to $\Delta E_2$ (the solid line in this figure).
It can be observed that both data analysis methods match nicely.

\begin{figure}
  \centering
  \subfigure{\includegraphics[width=.49\linewidth]{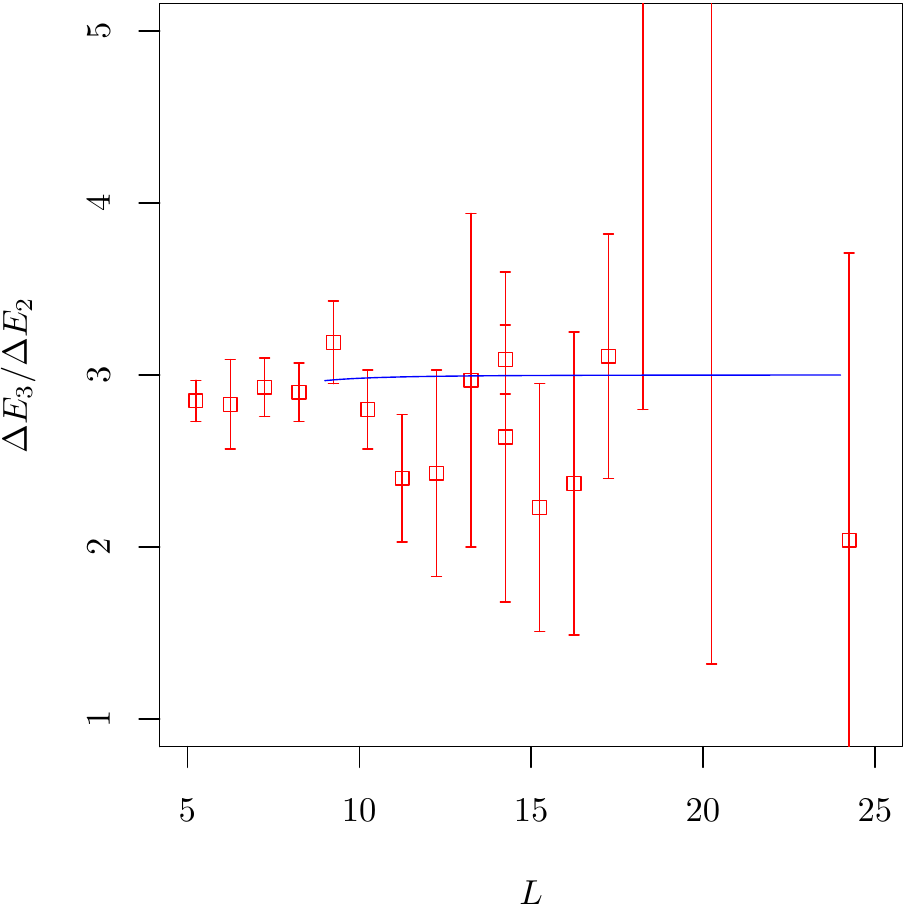}}
  \subfigure{\includegraphics[width=.49\linewidth]{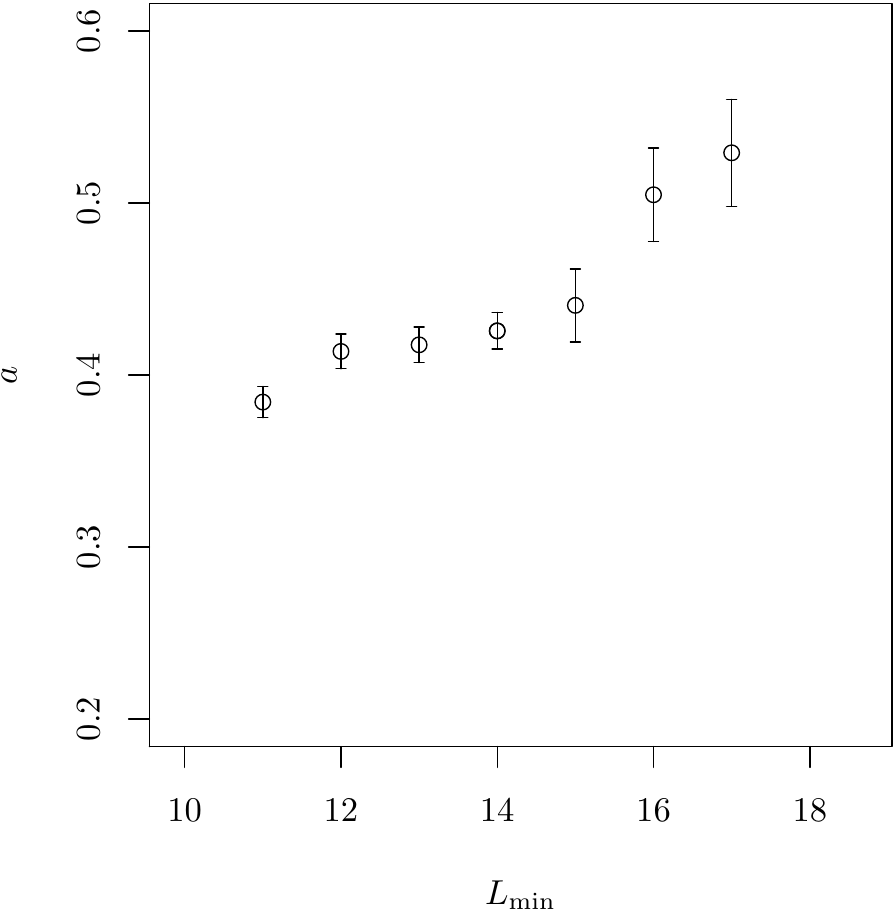}}
  \caption{Left: Ratio $\Delta E_3/\Delta E_2$ as a function of $L$.
    The solid line represents the ratio computed from the best fit parameters from Table~\ref{tab:selected}, using Eqs.~(\ref{eq:E2}) and (\ref{eq:E3}).
    Right: Scattering length $a$ as a function of $L_\mathrm{min}$
    determined fits of Eq.~(\ref{eq:E3}) up to order $L^{-5}$ to
    our data for $\Delta E_3$, i.e.
    with $a$ the only fit parameter.}
  \label{fig:DeltaE3-1}
\end{figure}

Turning to $\Delta E_3$ now, we have checked first that the data for
the ratio 
\begin{equation}
  \frac{\Delta E_3}{ \Delta E_2} =  3+ O(L^{-5}).
\end{equation}
is close to $3$, as one expects.
This is shown in the left panel of Fig.~\ref{fig:DeltaE3-1}.
Next we fit Eq.~(\ref{eq:E3}) to the
data for $\Delta E_3$, including only the scattering length $a$ as a single fit parameter.
Therefore, we include 
only the terms up to the order $L^{-5}$ in the fit.
If everything is consistent, this should allow us to reproduce the value of the
scattering length, determined by the fits to $\Delta E_2$ for sufficiently large values of
$L_\mathrm{min}$.

\begin{figure}
  \centering
  \subfigure{\includegraphics[width=.49\linewidth]{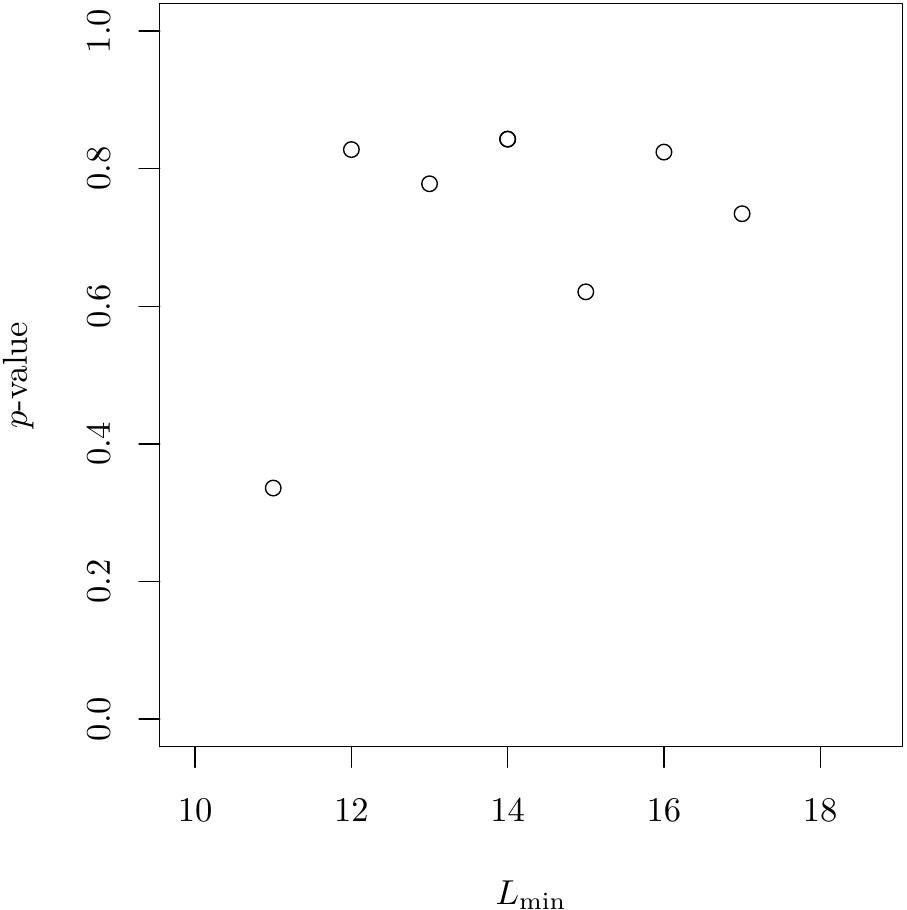}}
  \subfigure{\includegraphics[width=.49\linewidth]{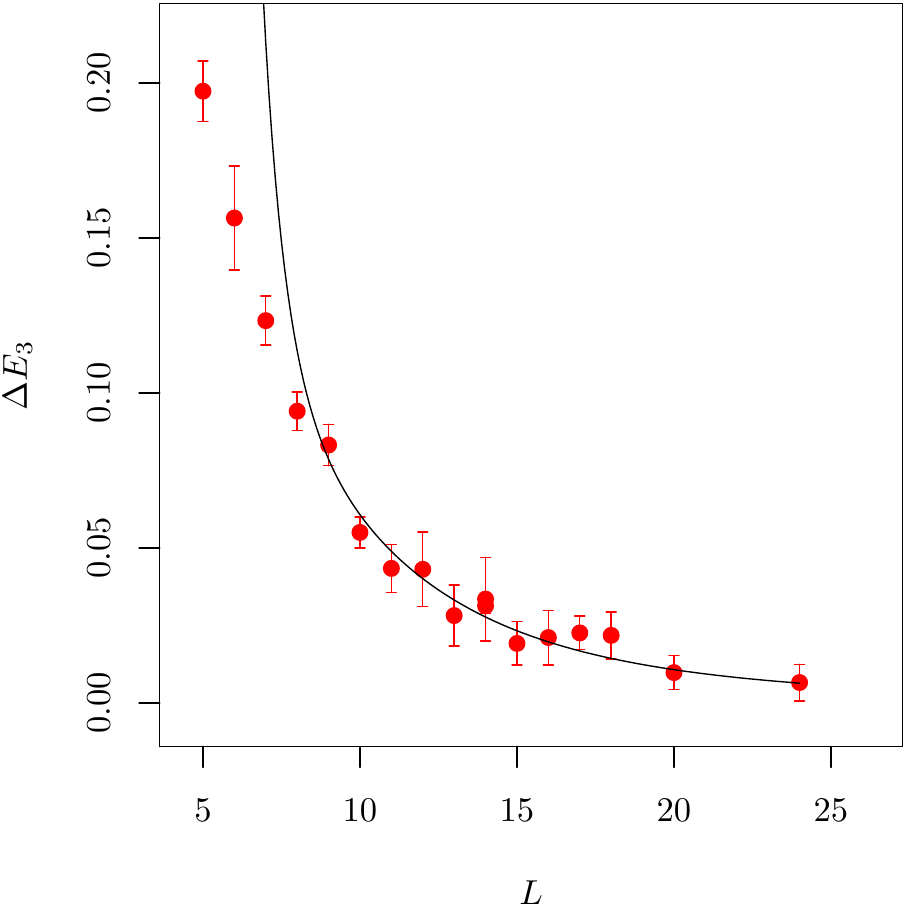}}
  \caption{Left: $p$-value as a function of $L_\mathrm{min}$
    for the fits of Eq.~(\ref{eq:E3}) up to order $L^{-5}$ to
    our data for $\Delta E_3$, i.e.
    with $a$ the only fit parameter.
    Right: Exemplary plot of a fit (solid line) of Eq.~(\ref{eq:E3})
    to the data for $\Delta E_3$ with $L_\mathrm{min}=9$ and
    $L_\mathrm{max}=24$.
    }
  \label{fig:DeltaE3fit}
\end{figure}

The result of this exercise is shown in the left panel of
Fig.~\ref{fig:DeltaE3-1} and the right panel of Fig.~\ref{fig:DeltaE3fit}.
We plot $a$ and the $p$-value of the fit,
both as functions of $L_\mathrm{min}$.
We observe reasonable $p$-values from $L_\mathrm{min}=11$ onward and the best
fit values for $a$ lie between $0.4$ and $0.5$. Within errors, this is
in agreement with the value quoted in Table~\ref{tab:selected}, which
was obtained
from the fit to $\Delta E_2$. 
We hence conclude that, for sufficiently large values of $L$,  the data for
$\Delta E_3$ and $\Delta E_2$ are reasonably well described by the same value
of the scattering length $a$.
We cannot perform the same test, including also $r$, because at that order also
the three-body term contributes.

As a next step, we will use the results for $a$ and $r$ from the fit of
Eq.~(\ref{eq:E2}) to the data for $\Delta E_2$ as priors for the fit of
Eq.~(\ref{eq:E3}) to the data for $\Delta E_3$, including terms up to the order $L^{-6}$.
To this end, we introduce an augmented $\chi^2$ function
\[
\chi^2_\mathrm{aug}\ =\ \chi^2 + \left(\frac{P_r - r}{\Delta r}\right)^2 + \left(\frac{P_a - a}{\Delta a}\right)^2
\]
with $\chi^2$ defined as usual, $P_r$ and $P_a$ being the priors for $a$ and $r$,
respectively, and $\Delta a$ and $\Delta r$ denoting the corresponding standard errors.
In the fit, we minimize the augmented $\chi^2_\mathrm{aug}$.
The actual values for $P_a, P_r, \Delta a$ and $\Delta r$ are the best fit parameters
with errors, which can be found in Table~\ref{tab:selected}.
The goal here is to see whether we get a significant contribution from the
three-body term in Eq.~(\ref{eq:E3}) or not.
An example of a fit with $L_\mathrm{min}=9$ and $L_\mathrm{max}=24$
is shown in the right panel of Fig.~\ref{fig:DeltaE3fit}.

\begin{figure}
  \centering
  \subfigure{\includegraphics[width=.49\linewidth]{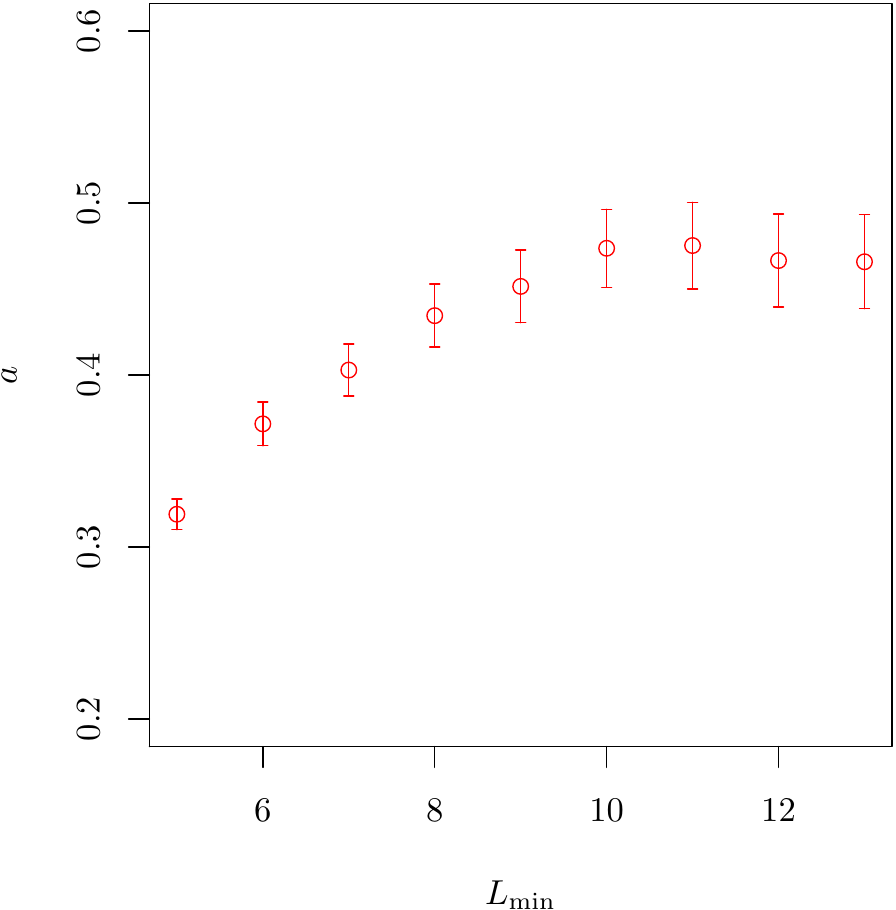}}
  \subfigure{\includegraphics[width=.49\linewidth]{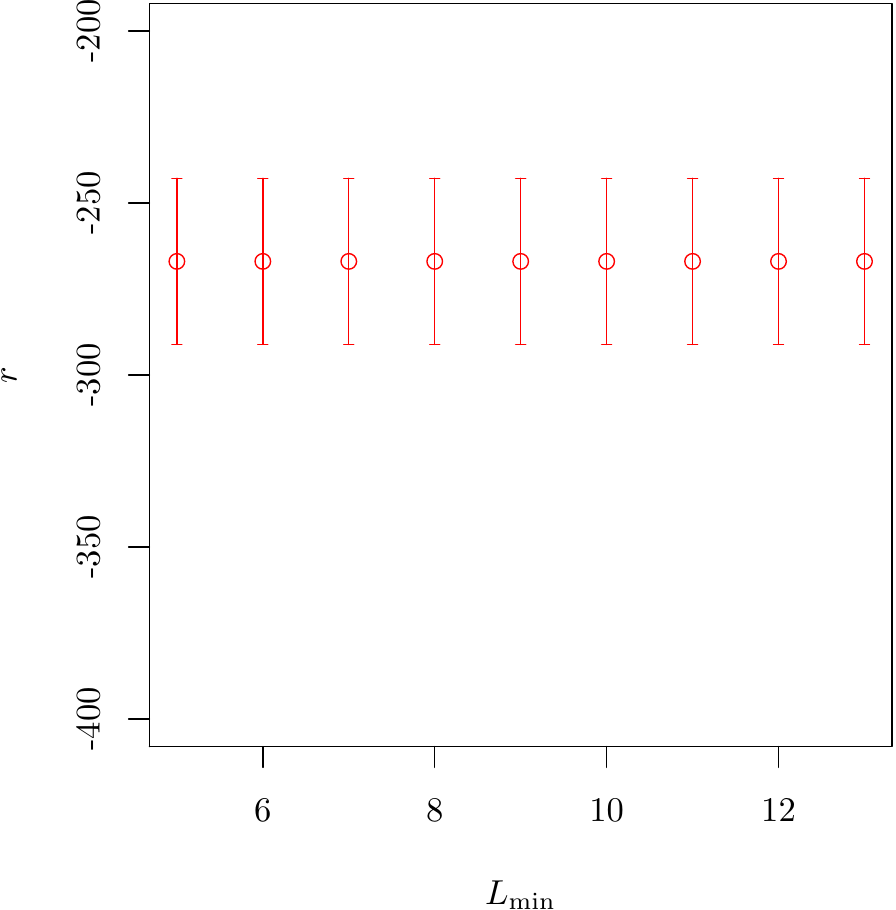}}
  \caption{$a$ and $r$ as functions of $L_\mathrm{min}$ from the fits of Eq.~(\ref{eq:E3}) to the data for $\Delta E_3$.
    Both $a$ and $r$ were included in the fit with priors from
    the fit to $\Delta E_2$ (the values can be found in Table~\ref{tab:selected}).}
  \label{fig:DeltaE3-2}
\end{figure}

The scattering length $a$ and the effective range $r$ are highly constrained by the prior included in the fit.
This can be observed from Fig.~\ref{fig:DeltaE3-2}, where $a$ (left panel) and $r$ (right panel) are plotted as functions of $L_\mathrm{min}$.
The fit to $\Delta E_3$ appears not to be sensitive to the effective range $r$, with
the prior value always reproduced.
This is expected, because the term proportional to $r$ interferes with the term
proportional to $D$.
Thus, the fit will always choose the prior value for $r$ to minimize the augmented $\chi^2$.
For $a$, we again observe that, for sufficiently large $L_\mathrm{min}$,
the value from the fit to $\Delta E_2$ is reproduced within errors.
From this, one may infer that $L_\mathrm{min}\geq 9$ should be chosen.

\begin{figure}
  \centering
  \subfigure{\includegraphics[width=.49\linewidth]{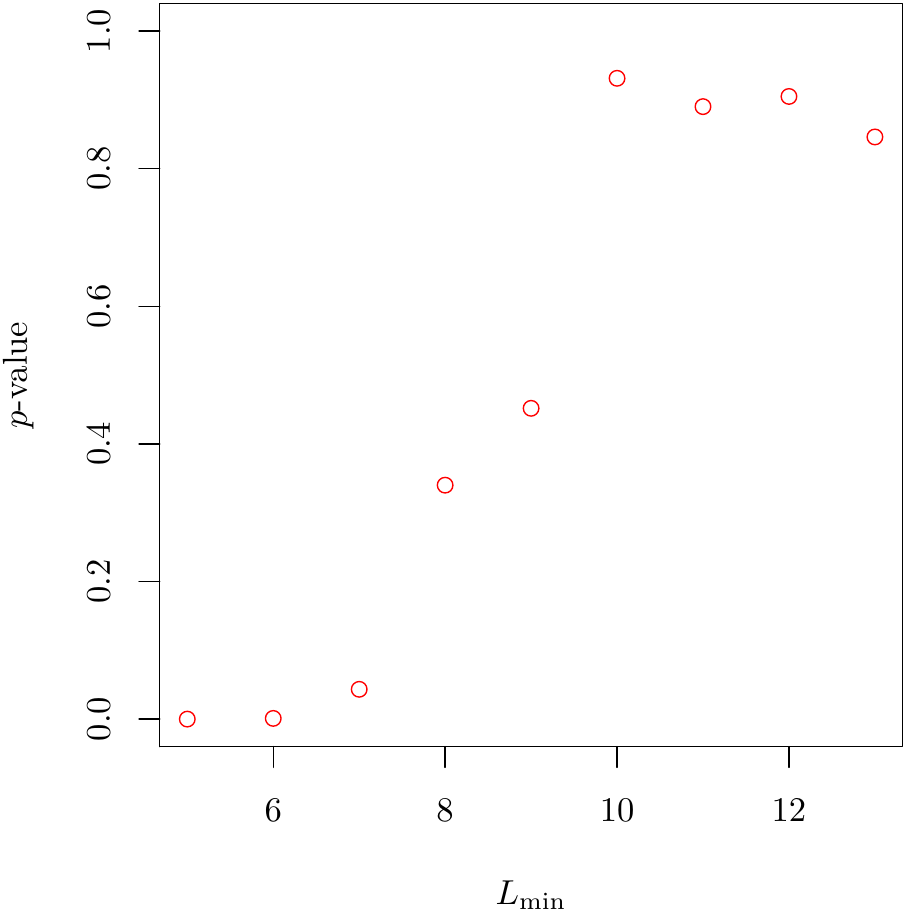}}
  \subfigure{\includegraphics[width=.49\linewidth]{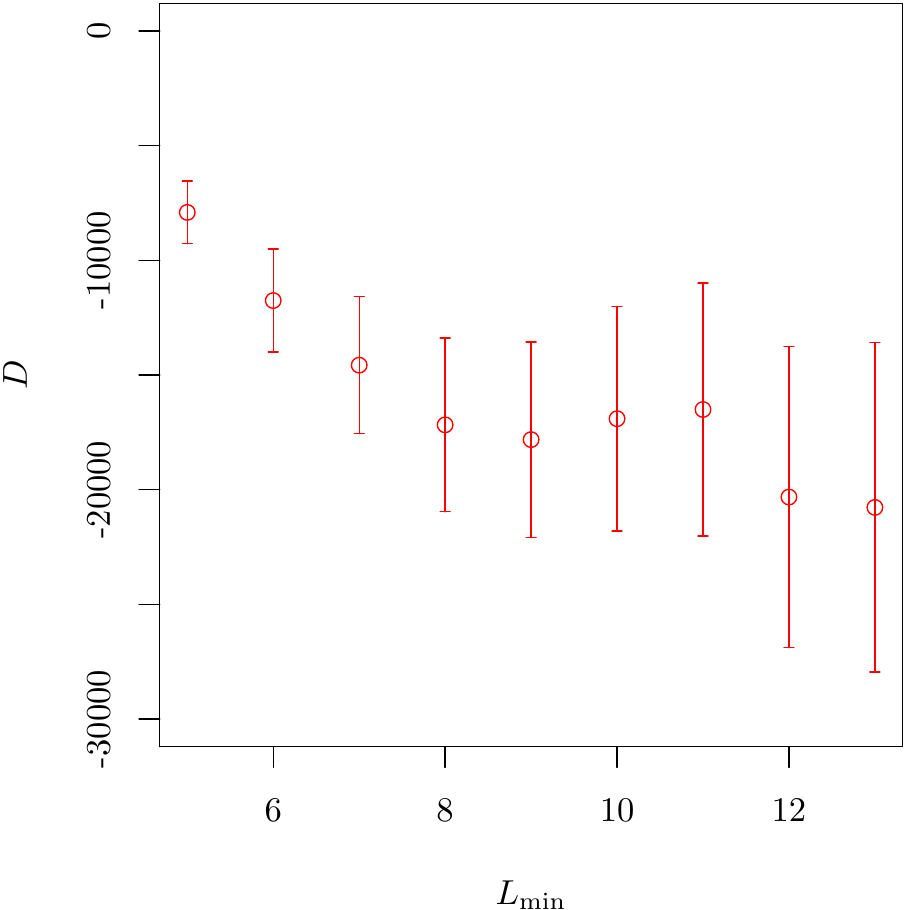}}
  \caption{$p$-value and $D$ as functions of $L_\mathrm{min}$ from fits of Eq.~(\ref{eq:E3}) to the data of $\Delta E_3$.}
  \label{fig:DeltaE3-3}
\end{figure}

This conclusion is also supported by the $p$-value, shown in the left panel of
Fig.~\ref{fig:DeltaE3-3} as a function of $L_\mathrm{min}$.
For $L_\mathrm{min}=8$ and $9$ we observe reasonable $p$-values.
For $L_\mathrm{min}=10$ or larger the $p$-values become close to $1$, indicating at
too many fit parameters for too few data points.
Finally, in the right panel of Fig.~\ref{fig:DeltaE3-3}, we show the three-body parameter $D$ as a function of $L_\mathrm{min}$.
In the relevant region of $L_\mathrm{min}$ we observe a plateau within errors
at a value significantly different from zero.
Actually, $D$ is different from zero to many $\sigma$ for all considered values of
$L_\mathrm{min}$, in particular $4\sigma$ for $L_\mathrm{min}=9$.
Final results of the fit to $\Delta E_3$ are shown in Table~\ref{tab:selected}.

\begin{table}[htpb!]
  \centering
  \begin{tabular*}{.8\textwidth}{@{\extracolsep{\fill}}ccrrrrr}
    \hline
    &$[L_\mathrm{min}, L_\mathrm{max}]$ & $a$         & $r$         & $D$  &$\chi^2/\mathrm{d.o.f.}$ & $p$-value \\ \hline
    $\Delta E_2$ &[9,24]       &0.46(3)      & -267(24)    & ---                   & 9.33/10          & 0.59     \\
    $\Delta E_3$ &[9,24]       &0.45(2)      & -267(24)    & -17814(4378)          & 10.90/11         & 0.45 \\ \hline
  \end{tabular*}
  \caption{Selected fit results to $\Delta E_{2,3}$}
  \label{tab:selected}
\end{table}

Eventually, we can use the parameters from Table~\ref{tab:selected}
to compute the ratio $\Delta E_3/\Delta E_2$, using Eqs.~(\ref{eq:E2}) and (\ref{eq:E3}),
both to the order $L^{-6}$.
The result is shown as a solid line in the left panel of Fig.~\ref{fig:DeltaE3-1}.
The agreement is satisfactory within error bars.

We close this section with two remarks:
First, we have also included an effective term of order $L^{-7}$ in the fits and observed that our results are stable under such a change.
Second, a direct fit to the data for $\Delta E_3/\Delta E_2$ appears not to be
sensitive to the fit parameters we are interested in.

\section{Summary and Outlook}

In this paper, we have investigated the two- and three-particle interactions in
complex $\varphi^4$ theory on the lattice.
In particular, we have carried out a theoretical study of the exponential finite-volume
corrections to the one-, two- and three-particle energy levels.
To next-to-leading order in perturbation theory it was shown that,
using the finite-volume particle mass $M_L$ instead of the infinite-volume
one $M$ in the definition of the two- and three-particle threshold energies
allows one to significantly reduce the exponential
corrections to the L\"uscher-type formulae for the two- and three-particle ground-state
energy shifts. For example,
in case of the model considered, these corrections get a suppression factor
$\lambda_c/L^2$, when $M_L$ is used. A proof of this statement to all orders
in perturbation theory seems challenging\footnote{Extending the proof to all orders
  for the Bethe-Salpeter kernel $K_L(p,p')$ by using the technique outlined in
  Ref.~\cite{Luescher-1} seems to be relatively straightforward. In order to complete the proof, however,
  one has to carefully analyze the exponential corrections, coming from the
two-particle propagator, to all orders as well.} but worth trying, since it could be
interesting for many lattice QCD applications. We plan to take up this challenge in
our future publications.

The big advantage of $\varphi^4$ theory as compared to QCD lies in a much reduced
computational complexity. This fact allowed us to study
a large set of ensembles with different finite volumes at fixed values of bare parameters.
At the first stage, we study the finite-volume corrections to the single-particle mass on
these ensembles.
We show that these corrections are well described by the corresponding expression known
from the effective field theory.
Next,  for each of the aforementioned ensembles, we determine the
two-particle and three-particle energy shifts in a finite volume,
$\Delta E_2$ and $\Delta E_3$.
Then, 
we extract the scattering length $a$ and the effective range $r$
from the $L$-dependence of $\Delta E_2$ by applying the L{\"u}scher formalism.
We also show that, depending on the range of values of $L$, used in the fit,
either only $a$ or $a$ and $r$ together can be determined, with consistent results for
$a$ in both cases.

As a next step, we show that
we can extract the scattering length for sufficiently large values
of $L$ from $\Delta E_3$ as well, and its 
value turns out to be consistent with the value extracted from the fit to $\Delta E_2$.
Hence, we conclude that our data for $\Delta E_2$ and $\Delta E_3$ are consistent.
Now, we follow the idea of Refs.~\cite{pang1,pang2} and use $a$ (and $r$),
determined from $\Delta E_2$, as an input for the analysis of $\Delta E_3$.
In this case, the (higher order) formula for  $\Delta E_3$ includes the three-body
effective coupling as well.
On the basis of the analysis of data, taken at many different volumes, we
find a statistically significant three-body contribution $D$, which is definitely away
from zero. This is a central result of our work\footnote{
{One should not be confused by the fact that
the three-body force emerges irrespectively to the absence of the $\varphi^6$
vertex in the Lagrangian. The notion of this force cannot be separated from the framework used to describe the energy levels in a finite box -- the non-relativistic effective field theory, in our case. In this theory, such a force is present from the beginning. In other words, we could simply state that our analysis is sensitive
to the terms of order $L ^{-6}$ in the three-particle energy, where the three-particle force starts to contribute.}}.

In this simple theory, there are still many issues to be explored. For example, one
could study the multiparticle states with more than three particles and try to
figure out, how their energy depends on the volume. Moreover, one may include excited three-particle states in the rest frame or moving frames for a more sophisticated study of the three-body coupling. In addition, it would be interesting
to investigate $\varphi^4$ theory at parameter values where a the
three-body bound state is present, since the volume-dependence of such
an energy level is
qualitatively different from the one seen it this work. For this, one would have to
further explore the parameter space of the theory.

\begin{acknowledgments}
  The authors thank R. Brice\~no, Z. Davoudi, M. Hansen and S. Sharpe
  for useful discussions.
 We acknowledge the support from the DFG (CRC 110 
``Symmetries and the Emergence of Structure in QCD'').
This research is supported in part by Volkswagenstiftung under contract no. 93562
and by Shota Rustaveli National Science Foundation (SRNSF), grant no. DI-2016-26. In addition, this project has also received funding from the European Union’s Horizon 2020 research and innovation
programme under the Marie Sk\l odowska-Curie grant agreement No. 713673. Special thanks to BCGS for their support.
\end{acknowledgments}

\begin{appendix}

\section{Spectrum \label{app:spectrum}}

\begin{table}[H]
  \centering
  \begin{tabular*}{.85\textwidth}{@{\extracolsep{\fill}}ccccccccc}
    \hline
    $L$& $T$ & $n_{conf}$ & $M_L$& $E_2(L)$& $E_3(L)$  & $\Delta E_2$
    & $\Delta E_3$& $\Delta E_3/\Delta E_2$     \\
    \hline
    4   & 24&18000 &0.3634(16)  & -- & -- & -- & -- & --  \\ 
    5   & 24&28000& 0.3049(13)  & 0.6790(20) & 1.1121(93) & 0.0692(24) & 0.1973(97) &2.85(12) \\
    6   & 24&7500& 0.2684(24)  & 0.5920(36) & 0.962(16) & 0.0552(46) & 0.156(17)&2.83(26)\\
    7   & 24&30000&0.2479(12)  & 0.5378(17) & 0.8669(74) & 0.0420(23) & 0.1233(79)&2.93(17) \\
    8   & 24&47000&0.2355(10)  & 0.5035(13) & 0.8006(57) & 0.0325(18) & 0.0941(62)&2.90(17) \\
    9   & 24&40000&0.2247(11)  & 0.4756(14) & 0.7574(62) & 0.0261(20) & 0.0832(67)&3.19(24) \\
    10  & 24&70000&0.21843(85) & 0.4565(11) & 0.7103(46) & 0.0196(15) & 0.0550(50)&2.80(23) \\
    11  & 24&30000&0.2142(13)  & 0.4464(17) & 0.6859(71) & 0.0181(23) & 0.0434(77)&2.40(37)\\
    12  & 24&12000&0.2095(21)  & 0.4367(26) & 0.672(11) & 0.0177(37) & 0.043(12)&2.43(60) \\
    13  & 24&20000&0.2088(16)  & 0.4271(21) & 0.6546(91) & 0.0095(28) & 0.0282(98)&2.97(97) \\
    14  & 24&28000&0.2054(22)  & 0.4236(28) & 0.650(13) & 0.0127(38) & 0.034(14)&2.64(96) \\
    15  & 24&40000&0.2057(12)  & 0.4199(15) & 0.6362(66) & 0.0086(20) & 0.0192(70)&2.23(72) \\
    16  & 24&52000&0.2045(14)  & 0.4179(18) & 0.6347(83) & 0.0089(25) & 0.0211(88)&2.37(88) \\
    17  & 24&70000&0.20540(87) & 0.4181(11) & 0.6388(50) & 0.0073(15) & 0.0226(54)&3.11(71) \\
    18  & 24&36000&0.2051(12)  & 0.4134(16) & 0.6371(71) & 0.0032(21) & 0.0218(76)&6.8(4.0) \\
    20  & 24&70000&0.20477(87) & 0.4114(11) & 0.6241(52) & 0.0018(15) & 0.0098(55)&5.4(4.1) \\
    14 &48&36000&0.20724(33) & 0.42461(63) & 0.6530(23) & 0.01014(62) & 0.0313(24)&3.09(20)\\
    24  &48&100000&0.20426(55) & 0.4118(11) & 0.6194(58) & 0.0032(10)
    & 0.0066(59)&2.0(1.7) \\
    \hline
  \end{tabular*}
  \caption{Energies, obtained from the lattice for one, two and three particles. Moreover, we include the energy difference and the ratio treating the correlations in a proper way. \label{tab:spectrum}}
\end{table}

\clearpage
\section{Effective mass and Correlation functions for $L=18$ \label{app:meff}}

\begin{figure}[H]
   \centering
   \subfigure[\ Effective mass for one particle at $L=18$.]{\includegraphics[width=0.475\textwidth,clip]{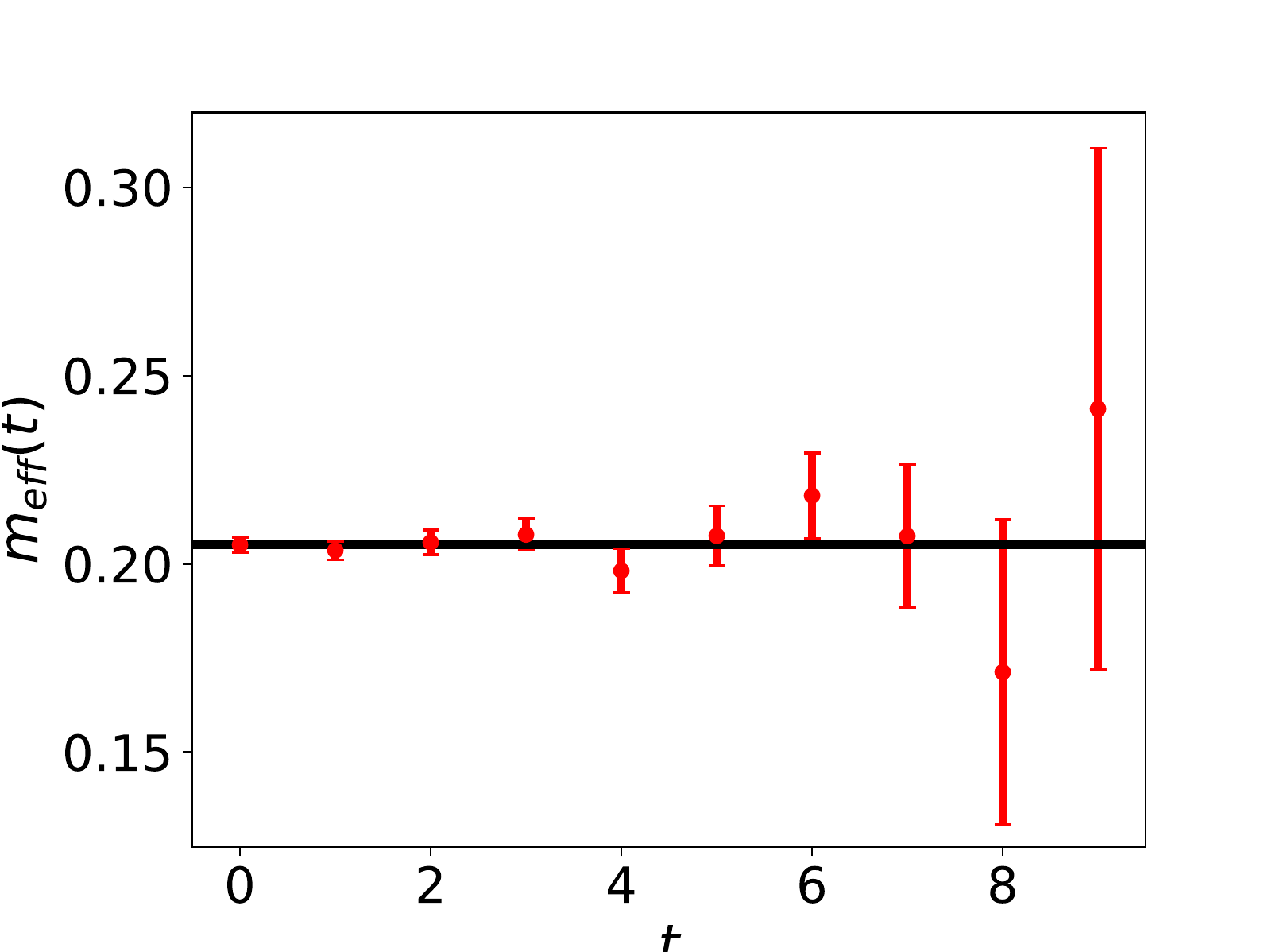}}\hfill   
   \subfigure[\ Correlation function of one particle for $L=18$]{\includegraphics[width=0.475\textwidth,clip]{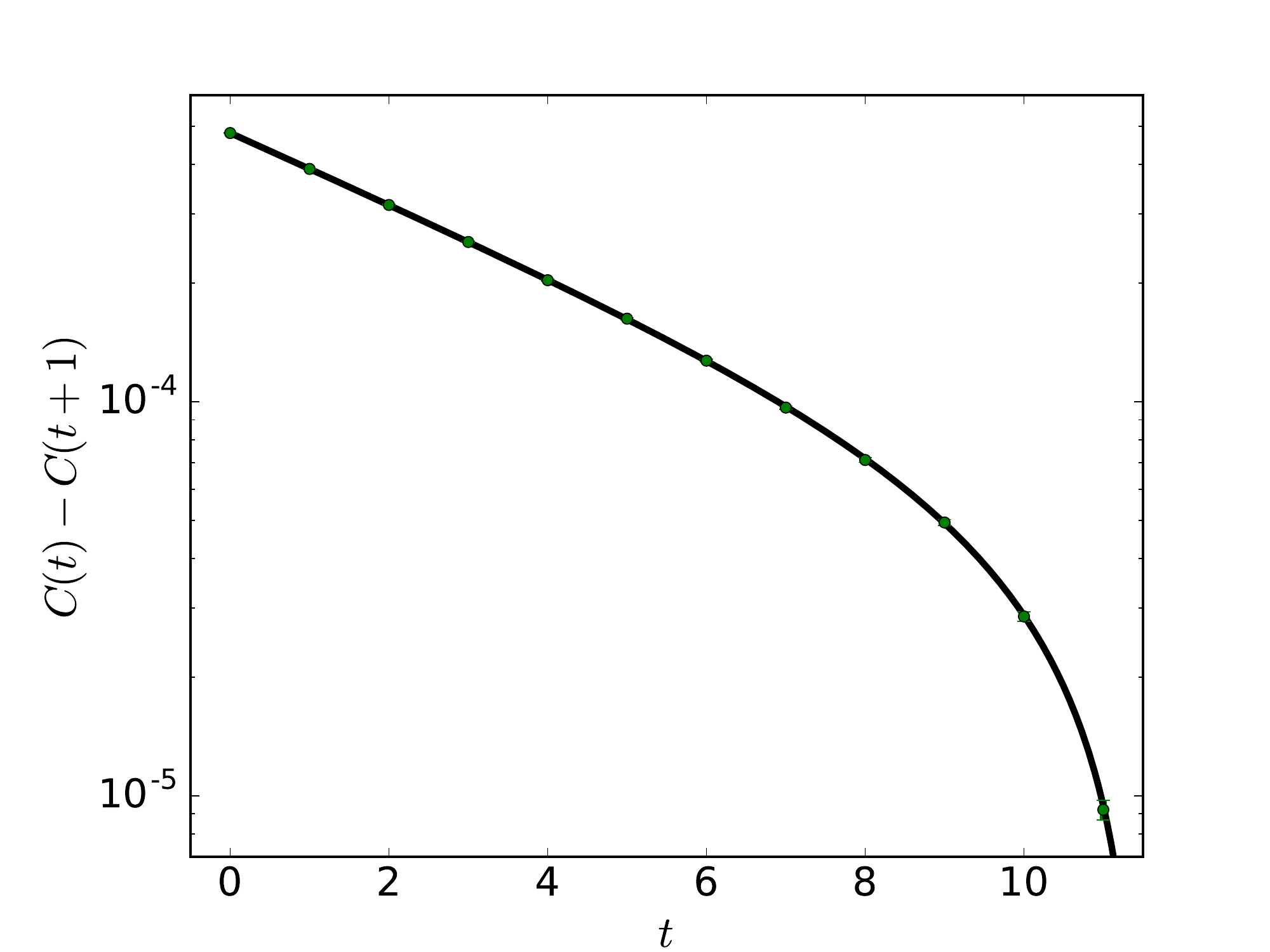}}  
   \subfigure[\ Effective mass for two particles in $L=18$.]{\includegraphics[width=0.475\textwidth,clip]{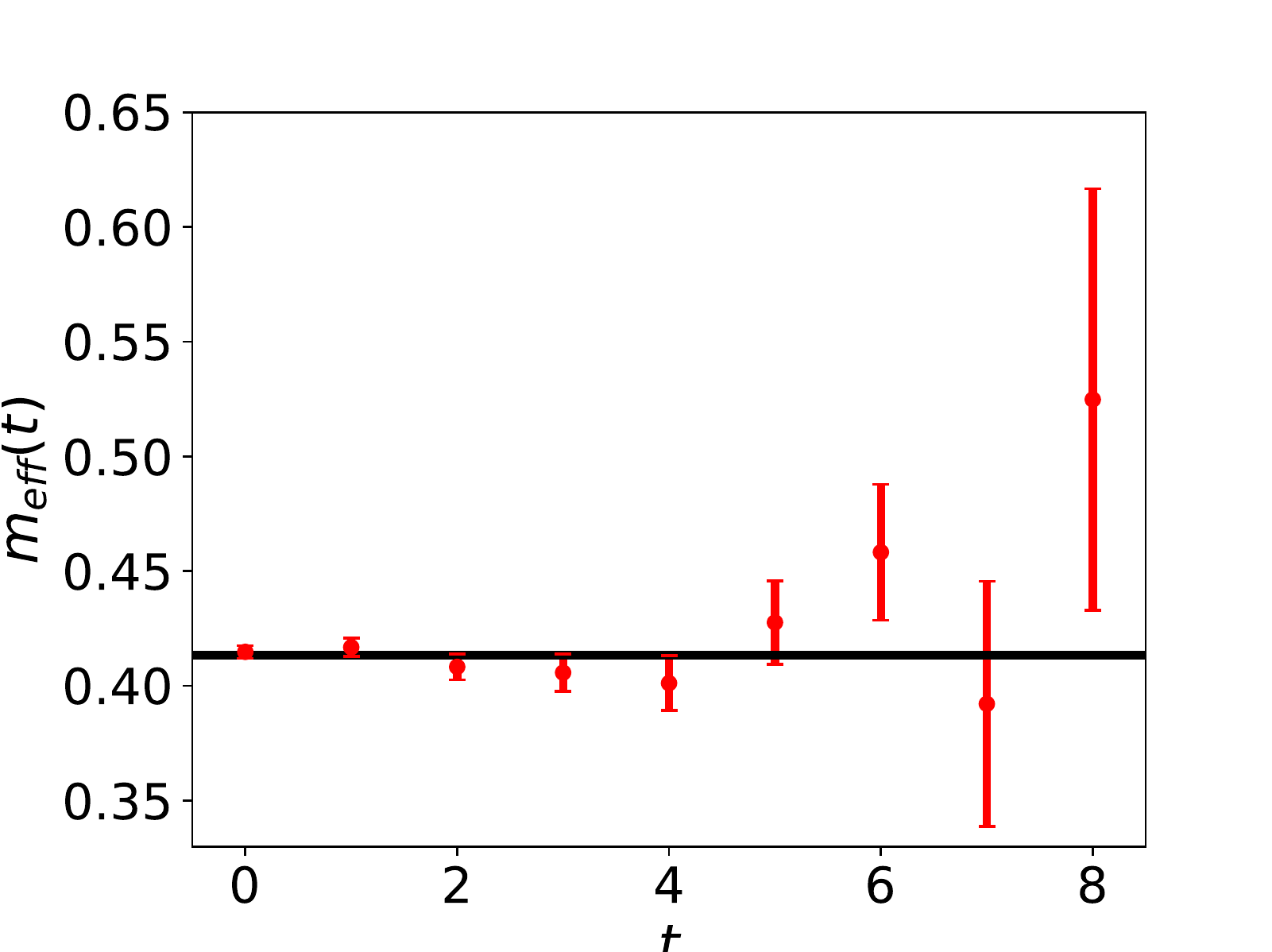}}\hfill   
   \subfigure[\ Correlation function of two particles for $L=18$]{\includegraphics[width=0.475\textwidth,clip]{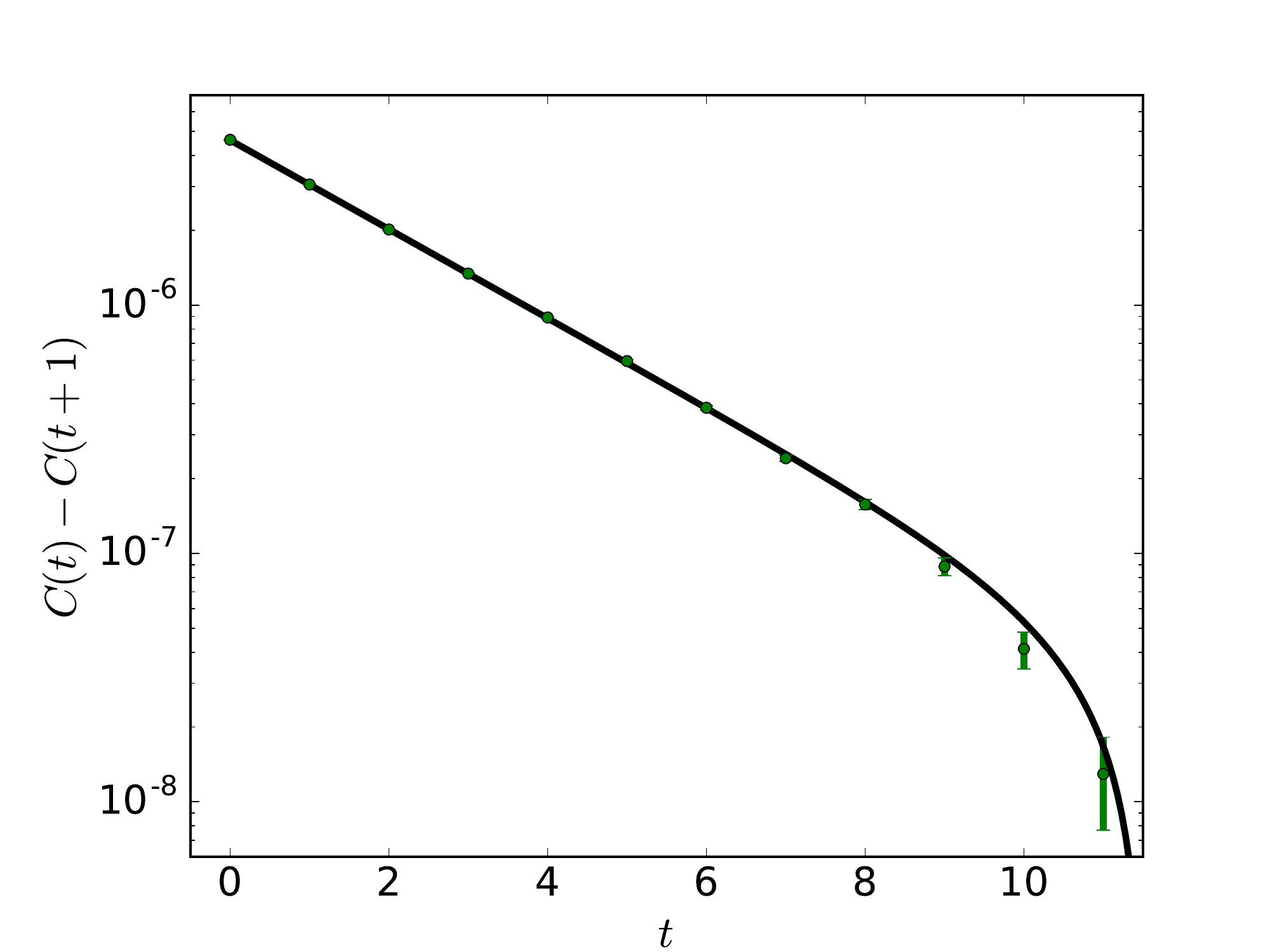}}  
   \subfigure[\ Effective mass for three particle in $L=18$.]{\includegraphics[width=0.475\textwidth,clip]{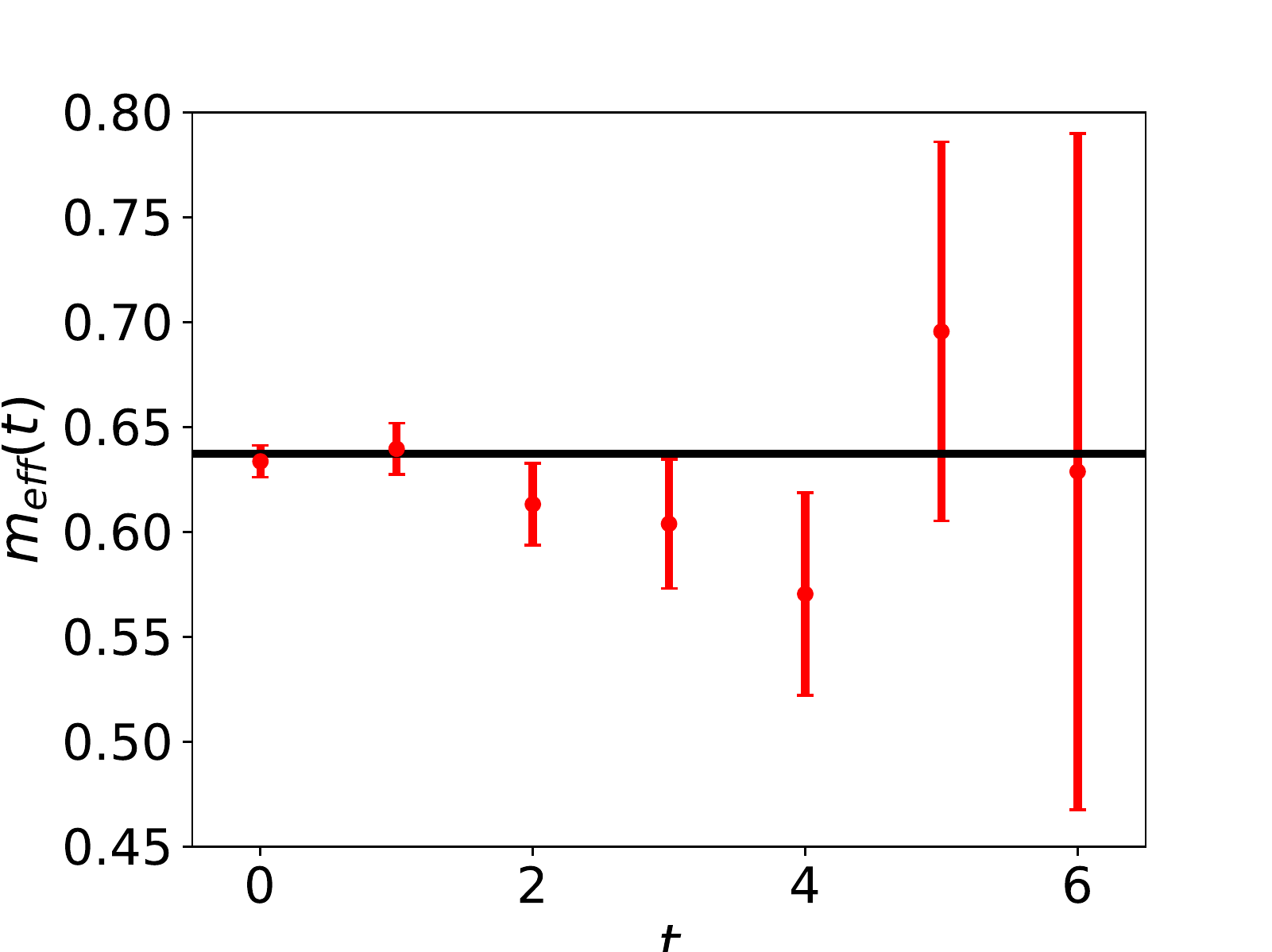}}\hfill   
   \subfigure[\ Correlation function of three particles for $L=18$]{\includegraphics[width=0.475\textwidth,clip]{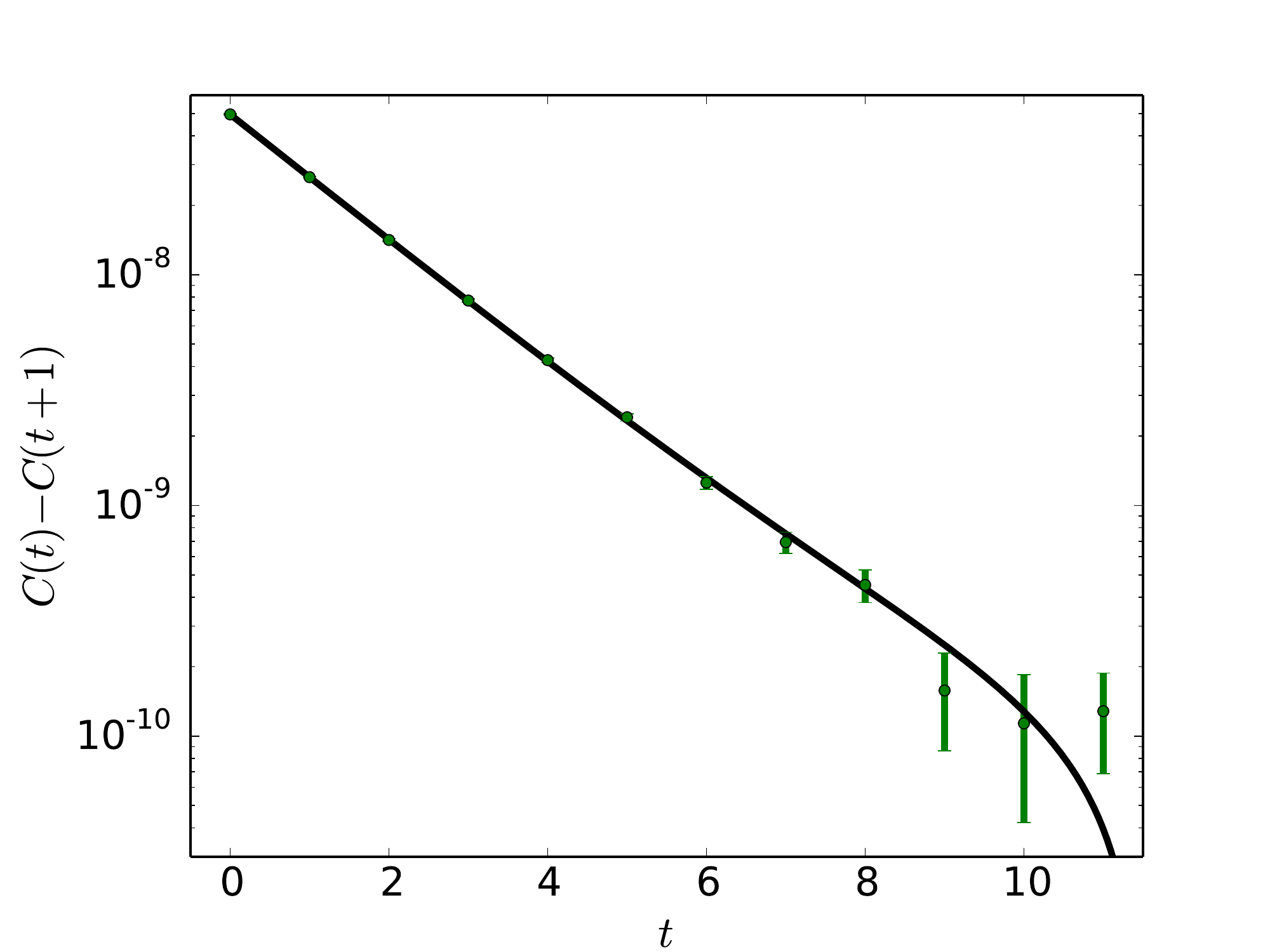}}  
   \label{fig:onepart}\caption{Relevant plots for our analysis at $L=18$ }
\end{figure}

\section{Phase Shifts \label{app:delta}}

\begin{table}[H]
  \centering
  \begin{tabular*}{.45\textwidth}{@{\extracolsep{\fill}}cccc}
    \hline
    $L$& $T$ & $k$           & ${\delta}$      \\
    \hline
    %4  &24& 0.1929(33) & -1.851(88)  \\ \hline
    5  &24& 0.1508(26) & -1.736(83)  \\ 
    6  &24& 0.1257(51) & -1.74(20)   \\ 
    7  &24& 0.1029(27) & -1.61(12)   \\ 
    8  &24& 0.0894(24) & -1.50(11)   \\ 
    9  &24& 0.0781(29) & -1.43(15)   \\ 
    10 &24& 0.0665(25) & -1.22(13)   \\ 
    11 &24& 0.0632(40) & -1.38(24)   \\ 
    12 &24& 0.0618(64) & -1.66(48)   \\ 
    13 &24& 0.0450(67) & -0.85(36)   \\ 
    14 &24& 0.0517(76) & -1.55(41)   \\ 
    15 &24& 0.0424(36) & -1.07(36)   \\ 
    16 &24& 0.0430(60) & -1.36(37)   \\ 
    17 &24& 0.0389(41) & -1.21(36)   \\ 
    18 &24& 0.0258(86) & -0.43(42)   \\ 
    20 &24& 0.0194(32) & -0.26(32)   \\ 
    24 &48& 0.0259(41) & -1.01(46)   \\ 
    14 &48& 0.0463(14) & -1.135(99)  \\
    \hline
  \end{tabular*}
  \caption{Phase shift in degrees, obtained from Eq.~(\ref{eq:delta}). The small value ensures perturbability. \label{tab:delta}}
\end{table}

\section{Fits $\Delta E_{2,3}$ \label{app:allfits}}

\begin{table}[H]
  \centering
  \begin{tabular*}{.6\textwidth}{@{\extracolsep{\fill}}ccccc}
    \hline
    $[L_i, L_f]$ & $a$         & $r$         &$\chi^2/\mathrm{d.o.f.}$
    & $p$-value \\
    \hline
    $[7,24]$       & 0.391(16)   & -191(10)    & 1.55  &  0.09 \\ 
    $[8,24]$       & 0.428(20)   & -225(14)    & 1.07  &  0.38 \\ 
    $[9,24]$       & 0.462(28)   & -267(24)    & 0.84  &  0.59 \\ 
    $[10,24]$      & 0.491(32)   & -309(32)    & 0.70  &  0.73 \\ 
    $[11,24]$      & 0.477(51)   & -270(180)   & 0.76  &  0.66 \\
    \hline
  \end{tabular*}
  \caption{Exemplary fit results for $\Delta E_{2}$. \label{tab:fitE2}}
\end{table}

\begin{table}[H]
  \centering
  \begin{tabular*}{.6\textwidth}{@{\extracolsep{\fill}}ccccc}
    \hline
    $[L_i, L_f]$ & $a$         & $D$         & $\chi^2/\mathrm{d.o.f.}$
    & $p$-value \\
    \hline
        $[7,24]$       & 0.403(15)   & -14569(3022)   & 1.75 & 0.04  \\ 
        $[8,24]$       & 0.435(18)   & -17169(3915)   & 1.12 & 0.34  \\ 
        $[9,24]$       & 0.452(21)   & -17814(4378)   & 0.99 & 0.45  \\ 
        $[10,24]$      & 0.474(23)   & -16903(5030)   & 0.43 & 0.93  \\ 
        $[11,24]$      & 0.475(25)   & -16502(5592)   & 0.48 & 0.89  \\
        \hline
  \end{tabular*}
  \caption{Exemplary fit results for $\Delta E_{3}$. \label{tab:fitE3}}
\end{table}

\end{appendix}

%\clearpage

%\tableofcontents
%\ordercite

\end{document}